\documentclass[fleqn,usenatbib,useAMS]{mnras}

\usepackage{graphicx}	
\usepackage{amsmath}	
\usepackage{amssymb}	
\usepackage{multicol}   
\usepackage[british]{babel}
\usepackage{csquotes}
\usepackage{changepage}
\usepackage{caption}
\usepackage[export]{adjustbox}
\usepackage{mwe}
\usepackage{float}
\usepackage{multirow}
\usepackage{bold-extra}
\usepackage{bm}	 
\usepackage{color}
\usepackage{pdflscape}
\usepackage{rotating}
\usepackage[table,xcdraw]{xcolor}

\usepackage[T1]{fontenc}
\usepackage{ae,aecompl}

\title[A deeper view of soft state of GRS 1915+105]{Decoding the Origin of HFQPOs of GRS 1915+105 during `Canonical' Soft States: An In-depth View using Multi-mission observations}

\author[P. Majumder et al.]{
Prajjwal Majumder$^{1}$\thanks{E-mail: \href{mailto:majumderprajjwal@gmail.com}
{majumderprajjwal@gmail.com}},
Broja G. Dutta$^{1}$\thanks{E-mail: \href{mailto:brojadutta@gmail.com}
{brojadutta@gmail.com}},
Anuj Nandi$^{2}$\thanks{E-mail: \href{mailto:anuj@ursc@gov.in}
{anuj@ursc.gov.in}}
\\
$^{1}$Department of Physics, Rishi Bankim Chandra College, Naihati, West Bengal 743165, India. \\
$^{2}$Space Astronomy Group, ISITE Campus, U. R. Rao Satellite Centre, Outer Ring Road, Marathahalli, Bangalore, 560037, India.
}

\date{Accepted XXX. Received YYY; in original form ZZZ}

\pubyear{2025}

\begin{document}
\label{firstpage}
\pagerange{\pageref{firstpage}--\pageref{lastpage}}
\maketitle

\begin{abstract}
We present a comprehensive analysis of the `canonical’ soft state ($\gamma$, $\delta$, and $\phi$ spectral variability classes) of the black hole binary GRS 1915+105, using \textit{RXTE}, \textit{AstroSat}, and \textit{NuSTAR} data from 1996 to 2017 to investigate the origin of High Frequency Quasi-periodic Oscillations (HFQPOs).
Our findings reveal that HFQPOs occur only in the $\gamma$ and $\delta$ classes, with frequencies of $65.07-71.38$ Hz and are absent in the $\phi$ class. We observe an evolution of time-lag from hard-lag (1.59$-$7.55 ms) in \textit{RXTE} to a soft-lag (0.49$-$1.68 ms) in \textit{AstroSat} observations.
Wide-band (0.7$-$50 keV) spectral modelling suggests that HFQPOs are likely observed with a higher covering fraction ($f_{cov} \gtrsim 0.5$), i.e., the fraction of seed photons being Comptonized in the corona, enhanced Comptonized flux ($\sim$ 38\%), and lower optical depth ($\tau \lesssim 8.5$ ) in contrast to observations where HFQPOs are absent. We observed similar constraints for observing HFQPOs   
during an inter-class ($\phi \rightarrow \delta$) transition
as well as in a few intra-class ($\delta \rightarrow \delta$) variations.
We also find that the time lag decreases as $\tau$ increases, indicating that a higher $\tau$ reduces Compton up-scattering, thereby decreasing the hard-lag.
Interestingly, in \textit{RXTE} observations, the hard-lag ($\sim$ 7 ms) gradually decreases as optical depth and Comptonization ratio increases, eventually becoming a soft-lag ($\sim$ 1 ms) in \textit{AstroSat} observations. 
These constraints on spectro-temporal parameters for the likelihood of observing HFQPOs support a `compact' coronal oscillation mechanism for generating HFQPOs, which we attempt to explain within the framework of a possible accretion scenario.

\end{abstract}




\section{Introduction}

Black-hole X-ray binaries (BH-XRBs) are systems consisting of black holes that accrete matter from a companion star and emit significant X-ray radiation in the process. These systems provide valuable insights into extreme gravitational environments and serve as laboratories for testing general relativity \citep{tregidga_2024}. 
The behaviour of BH-XRBs is complex and influenced by several factors such as the mass, spin of the black hole, inclination angle etc. Their X-ray luminosity can vary significantly, ranging from quiescent states with luminosity as low as a few $10^{30}$ erg/s \citep{gallo_low_lumin-bh_2008}, to as high as $\sim$ $10^{41}$ erg/s in ultraluminous X-ray sources \citep{fabbiano_ulx_1989,feng_soria_xmm_ulx2011,seshadri_ulx_2023}.

It is well known that a typical BH-XRB progresses through a number of `canonical' states during an outburst \citep[][and references therein]{homan_bh-state2001,remillard_mcClintock2006,anuj_nandi_gx339_2012,sreehari2019,anuj_nandi2024}. The transient nature of many BH-XRBs is attributed to thermal-viscous instabilities in their accretion discs \citep{lasota_bh-transient_2001}. Depending on the hardness and intensity variation of the source, all observations can be divided into four `canonical' states: low hard state (LHS), hard intermediate state (HIMS), soft intermediate state (SIMS) and high soft state (HSS) \citep{homan_bh-state2001,remillard_mcClintock2006,anuj_nandi_gx339_2012}. The different spectral states reflect various configurations of the accretion flow dynamics and associated dominating emission mechanisms.
Spectral modelling of various BH-XRBs indicates the presence of a multi-temperature Keplerian accretion disc \citep{shakura_sunyaev1973} that contributes to the thermal emission of the source. Additionally, the high-energy non-thermal emission is thought to have originated in the `hot' corona near the black hole due to the inverse Compton scattering of seed blackbody photons by high energetic electrons \citep{sunyaev_titarchuk1980,chakrabarti_titarchuk1995,zdziarski1996}. During the hard state observations, the optically thick and geometrically thin disc is truncated at the very large radii and the whole spectrum is dominated by the high-energy non-thermal emission. As the disc moves progressively inward, the thermal emission starts to dominate the spectra and the source is observed to be in the soft state \citep{skc_dutta_pal_2009,broja_skc_2010,iyer_nandi2015,dutta2016,dutta2018}. The inner radius of the accretion disc is observed to be close to the innermost stable orbit of the black hole during HSS. 
Thus, HSS observations of BH-XRBs could shed light on a better understanding of the accretion processes and energy transfer in close proximity to the event horizon, where relativistic effects are dominant.

The X-ray spectra in the soft state exhibit notable reflection characteristics of iron K$\alpha$ emission lines which is believed to be resulted from X-rays emitted by the corona illuminating the accretion disc \citep{ross_fabian_refl_2007}.
X-ray reflection spectroscopy is a powerful tool for measuring black hole spins \citep{reynolds_bh-spin_2014} and testing general relativity in strong gravitational fields \citep{bambi_2017}. 
Modelling the soft state in BH-XRBs is challenging due to complex processes near black holes, particularly in accurately representing the inner accretion disc, where relativistic and gravitational effects are strongest.
However, polarization studies can shed a light to decipher the accretion geometry and radiation mechanisms during the soft state. Interestingly, \textit{IXPE} observations have detected a high polarization of $\sim$ 8\% during the soft state of 4U 1630-47 \citep{rawat_2023a_4U1630, kushwaha_2023, ratheesh_ixpe_2024}. In contrast, relatively low polarization values of 3\% and 2\% have been observed in the soft states of LMC X-3, Cyg X-1, and 4U 1957+115, respectively \citep{seshadri_2024_lmcx3,marra_2024,steiner_ixpe_cygx1_2024}. Thus, even with polarization observations, understanding and explaining the nature of soft state of a BH-XRBs remains challenging, as the polarization fraction varies across the sources for the similar spectral state, indicating crucial role of differences in accretion geometry and radiation mechanisms.

HFQPOs are typically observed during the soft or soft-intermediate states of a source when disc emission is predominant \citep{belloni2012}. 
HFQPOs are thought to be manifestations of various relativistic effects in orbits near black holes, making them important tools for investigating general relativity in extreme gravitational conditions \citep[][and references therein]{stella_vietri1998, rebusco2008, merloni1999, vincent2013, stefanov2014}. Nevertheless, a definitive explanation for the origin of HFQPOs remains elusive. \cite{cui1999} proposed that HFQPOs ($\sim$ 67 Hz) to be coupled with Comptonizing region based on significant hard lags in GRS 1915+105, while \cite{remillard_2002} emphasized the role of a `Compton corona' in reprocessing disk photons. \cite{aktar2017,aktar2018} attributed the 300 Hz and 450 Hz oscillations in GRO J1655-40 due to modulations in the post-shock corona. \cite{dihingia2019} indicated that shock-induced relativistic accretion solutions could potentially explain the oscillations in well-studied sources such as GRS 1915+105 and GRO J1655-40.
These HFQPOs are particularly important due to their relatively stable centroid frequency, which remains almost unaffected by significant changes in luminosity. This characteristic distinguishes HFQPOs in black hole binaries from the variable kHz QPOs found in neutron stars \citep{remillard_mcClintock2006}.
HFQPOs are detected in few BH-XRBs observed with \textit{RXTE}, such as, GRS 1915+105, GRO J1655-40, XTE J1550-564, XTE J1859+226, H 1743-322, 4U 1630-47 and IGR J17091-3624 \citep[][and references therein]{belloni2012,altamirano2012,belloni2013}. GRS 1915+105 exhibits a generic HFQPO $\sim$ 67 Hz persistently for over the last 25 year \citep{morgan1997,belloni2013,belloni2019,sreehari2019,seshadri2022}.

GRS 1915+105, a microquasar \citep{mirabel1994} is a very bright BH-XRB source that was first discovered by WATCH in 1992. The source consists of a K-M III type companion star with mass $1.4M_{\sun}$ with an orbital period of 33.5 days \citep{greiner2001} and a black hole of mass $12.4M_{\sun}$ at a distance of $8.6$ kpc \citep{reid2014}. The black hole possibly has a spin > 0.98 measured indirectly \citep[see][and references therein]{sreehari2020}. Unlike any other black hole sources, GRS 1915+105 exhibits different types of variability in its lightcurve with a time scale of minutes to seconds. Depending on its structure of lightcurve and colour-colour diagram all observations are classified into 14 distinct classes \citep{belloni2000,klein-wolt2002,hannikainen2005}. This source has been persistently bright over 25 years and showed all the spectral and timing features throughout these years. 
HFQPOs in the 63$-$71 Hz range have been observed across seven variability classes ($\kappa$, $\gamma$, $\mu$, $\delta$, $\omega$, $\rho$, and $\nu$) associated with the soft or soft-intermediate state using \textit{RXTE} \citep{belloni2013} and in four variability classes ($\delta$, $\kappa$, $\omega$ and $\gamma$) using \textit{AstroSat} \citep{seshadri2022} observations.
In this study, we focus on differentiating spectro-temporal quantities specifically during the canonical soft state, regardless of HFQPO presence. Therefore, we have selected \textit{RXTE}, \textit{AstroSat}, and \textit{NuSTAR} observations of the $\gamma$, $\delta$, and $\phi$ classes, as these variability classes correspond to the canonical soft state.

In the \textit{AstroSat} era, we performed extensive time-lag studies \citep{prajjwal_2024} for the $\delta$, $\omega$, $\kappa$, $\gamma$ variability classes and found soft-lag with respect to $3-6$ keV energy band, whereas \cite{belloni2019} observed hard-lag with respect to the $5-10$ keV band. However, hard-lag were observed during \textit{RXTE} observations \citep{cui1999,mendez2013}.
Furthermore, \cite{seshadri2022} conducted an extensive analysis of all \textit{AstroSat} observations and concluded that the HFQPOs are significant in the $6-25$ keV band and appear only in the $\delta$, $\omega$, $\kappa$, $\gamma$ variability classes. 
\cite{sreehari2020, seshadri2022} suggested, based on \textit{AstroSat} observations, that in the soft spectral state, the HFQPOs may result from modulation of the Comptonizing corona near the source.
\cite{ueda_2009} analysed this source during the soft state with \textit{RXTE} and found evidence of both Comptonization and reflection feature in the \textit{PCA} spectra.
Interestingly, a long-term correlated analysis of HFQPOs during soft spectral state, using observations from both \textit{AstroSat} and \textit{RXTE}, has not yet been conducted by anyone. 
Therefore, exploring the spectro-temporal characteristics of this source in the soft state could provide valuable insights into the origin of HFQPOs.
In this paper, we conduct an in-depth comprehensive study of all the `canonical' soft state observations (316 ks) of GRS 1915+105 using data from \textit{RXTE}, \textit{AstroSat}, and \textit{NuSTAR}. We have considered all observations of $\gamma$, $\delta$, and $\phi$ classes spanning from 1996 to 2017.

We have organized this paper as follows. In $\S$\ref{sec:obs_data_reduction}, we present the reduction procedures for \textit{RXTE}, \textit{AstroSat}, and \textit{NuSTAR} data. In $\S$\ref{sec:timing}, we discuss the timing analysis methods and present the results. The procedure for wide-band spectral analysis and the results obtained for all observations are discussed in $\S$\ref{subsec:spectral_analysis}. The correlation between spectral and temporal parameters is explored in $\S$\ref{sec:spectro-temporal_corr}. In $\S$\ref{sec:discussion}, we discuss the results in detail from the spectro-temporal analysis and address the origin of HFQPOs and the possible underlying accretion dynamics in soft spectral state. Finally, we conclude in $\S$\ref{sec:conclusion}.

\begin{figure*}
 \includegraphics[width=0.95\textwidth]{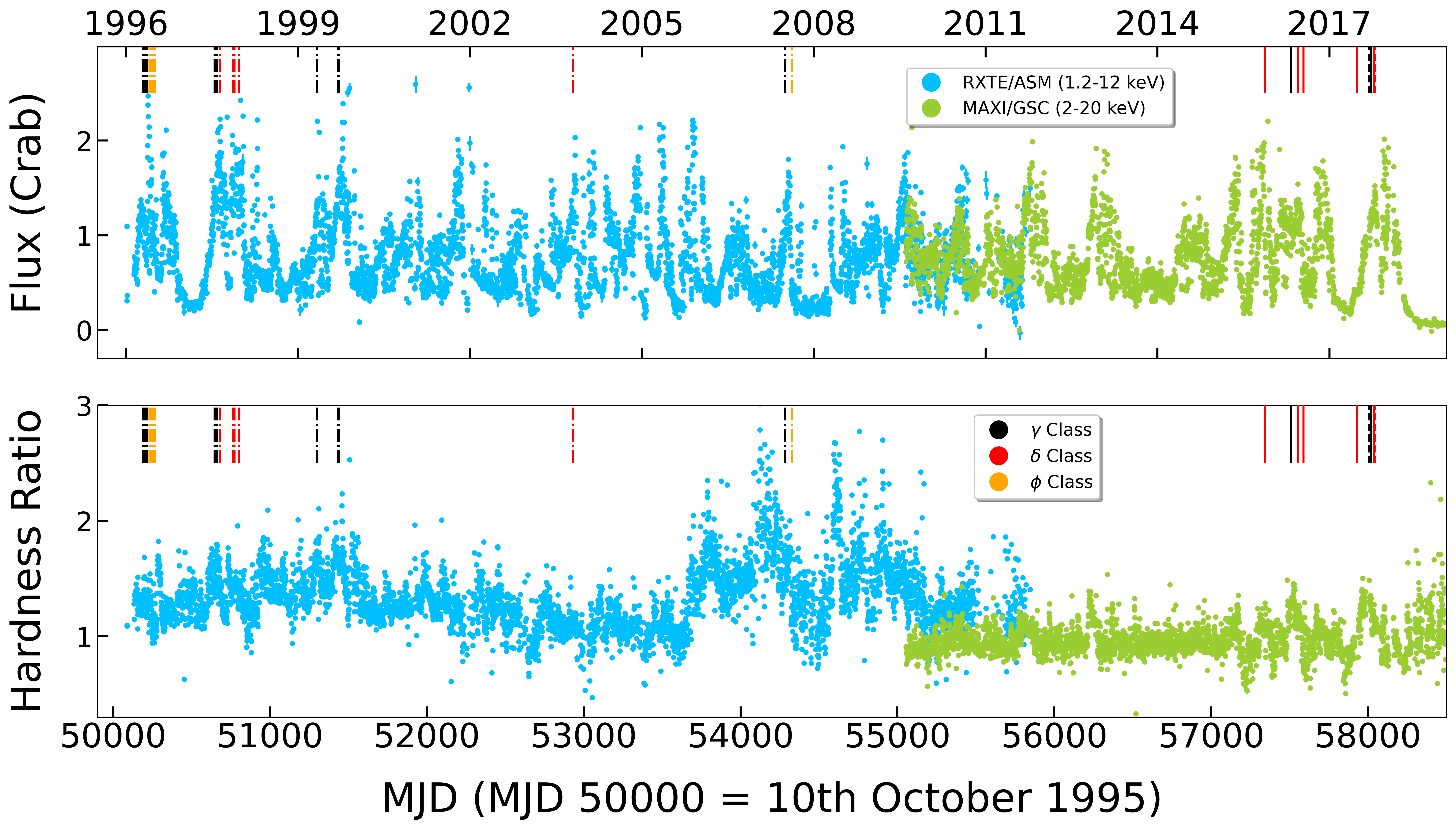}
 \caption{The \textit{RXTE/ASM} ($1.2-12$ keV, in blue) and \textit{MAXI/GSC} ($2-20$ keV, in green) light curves of GRS 1915+105 from 1996 to 2019 are plotted in the upper panel, with the hardness ratio (HR) shown in the lower panel. The \textit{RXTE/ASM} hardness is defined as, HR $= (5-12)$ keV$/(3-5)$ keV and the \textit{MAXI/GSC} hardness is defined as, HR $=(4-10)$ keV$/(2-4)$ keV. The vertical lines represent the observations considered in this work. The black, red and orange vertical lines correspond to the $\gamma$, $\delta$ and $\phi$ class observations, respectively. The \textit{RXTE}, \textit{AstroSat} and \textit{NuSTAR} observations are denoted by `dash-dot', `dashed' and `solid' lines, respectively. See text for details.}
 \label{fig:asm_maxi}
\end{figure*}

\section{Observation and Data Reduction}
\label{sec:obs_data_reduction}

We analysed observations consisting of three variability classes: $\gamma$ (18 observations), $\delta$ (26 observations), and $\phi$ (6 observations) during its `canonical' soft state \citep{belloni2000}, using data from \textit{RXTE}, \textit{AstroSat}, and \textit{NuSTAR} spanning from April 17, 1996 (MJD 50190.578) to October 19, 2017 (MJD 58045.931). The long-term \textit{RXTE/ASM} and \textit{MAXI/GSC} lightcurve of the source is plotted, with the considered observations marked as vertical line in Fig. \ref{fig:asm_maxi}.
In this work, we considered total 50 observations with a total exposure time of 316 ks. The list of all observations, with the missions,  MJD start and effective exposure are mentioned in Table \ref{tab:log_table}.

\subsection{\textit{RXTE} Data Reduction}
\textit{RXTE} has observed GRS 1915+105 through All-Sky Monitor (\textit{ASM}) \citep{levine_rxte/asm_1996} extensively from 1996 to 2012. We obtain the \textit{ASM} data from the HEASARC website\footnote{\label{web:heasarc_asm}\url{https://heasarc.gsfc.nasa.gov/docs/xte/asm_products.html}} to examine the one-day averaged lightcurve of the source in the $1.2-12$ keV band and its hardness ratio, defined as the ratio of count rates in $5-12$ keV to $3-5$ keV. We converted the full energy band \textit{ASM} count rate to Crab units by dividing each count rate by 75 counts/s \citep{punsly2013,motta2021}. 
The long-term lightcurve and the hardness ratio are shown in Fig. \ref{fig:asm_maxi}. The \textit{RXTE/PCA} observations during this period are denoted by the vertical coloured `dash-dot' line. The colours are explained in the inset in the lower panel.

The \textit{RXTE} archival observations are obtained from the HEASARC website\footnote{\label{web:heasarc}\url{https://heasarc.gsfc.nasa.gov/cgi-bin/W3Browse/w3browse.pl}}.
The timing analysis of \textit{PCA} \citep{jahoda_2006_pca} data is performed in the energy range $2-60$ keV.
In order to carry out the dead-time corrected spectral analysis, we have used the ftool \texttt{pcaprepobsid} and \texttt{pcaextspect2} available in \texttt{HEASOFT v6.29c}\footnote{\url{https://heasarc.gsfc.nasa.gov/lheasoft/download.html}}, which produces the dead-time corrected source spectra, background spectra, and the response file for further spectral analysis\footnote{\url{https://heasarc.gsfc.nasa.gov/docs/xte/recipes2/Overview.html}}.  
\textit{HEXTE} \citep{rothschild_hexte_1998} is used only for the spectral analysis of all the observations. Among the two clusters of \textit{HEXTE}, the cluster A data is used. 
Background files could not be produced for Obs. 5, 19, 20, 21, 45, 46, 47, 48 and 49 (see Table \ref{tab:log_table}). Hence, the \textit{HEXTE} spectra for these observations could not be considered.
We generated wide-band (3$-$50 keV) energy spectra from \textit{RXTE} combining \textit{PCA} (3$-$30 keV) and \textit{HEXTE} (20$-$50 keV). A systematic error of 0.5\% is added for both \textit{PCA} and \textit{HEXTE} spectra \citep{shaposhnikov_rxte_systematic2012}.

\subsection{\textit{AstroSat} Data Reduction}

\textit{AstroSat} \citep{agrawal2006} is the first multi-wavelength satellite launched by India in 2015. We have used \textit{SXT} \citep{singh2017} and \textit{LAXPC} \citep{yadav2016,antia2017} observations for our spectro-temporal analysis of the source.
\textit{SXT} observes the X-ray sources in the energy range 0.3$-$8 keV with a timing resolution of 2.3775 s.
We have performed only the spectral analysis of level-2 data from \textit{SXT} available at \textit{AstroSat} public archive\footnote{\label{fn3}\url{https://webapps.issdc.gov.in/astro_archive/archive/Home.jsp}} due to its poor time resolution. We selected a 12 arcmin circular region for PC mode observations, while for FW mode observations, we used a 5 arcmin circular region.
The detailed data reduction for \textit{SXT} is carried out following \cite{seshadri2022,prajjwal_2024}.

We used \textit{LAXPC} level-1 data available in \textit{AstroSat} public archieve$^{\ref{fn3}}$ for both timing and spectral analysis. For timing analysis, \texttt{LAXPCsoftware} is used to process the level-1 data to level-2 data. We also used top-layer single event spectra, as recommended by \cite{antia2021}, for sources in the soft state to minimize the undesired spectral residual bump at $\sim$ 33 keV, corresponding to the Xenon K-edge \citep{antia2017, sreehari2019}.
However, top layer single event \textit{LAXPC20} spectra are extracted using \texttt{LaxpcSoftv3.4} \citep{antia2017} for our spectral analysis.

Wide-band (0.7$-$50 keV) spectral analysis was performed for \textit{AstroSat} observations combining \textit{SXT} and \textit{LAXPC} in the energy range 0.7$-$7 keV and 3$-$50 keV respectively. \textit{SXT} spectra are grouped with 30 counts in each bin, whereas grouping is not applied for \textit{LAXPC20} spectra. We added 2\% systematic error for both spectra \citep{antia2017}. In order to fit the instrumental edges at 1.8 keV and 2.2 keV due to Si and Au \citep{singh2017}, we added \texttt{gain fit} command in \texttt{XSPEC V12.12.0} \citep[see][and references therein]{prajjwal_2024}.

\subsection{\textit{NuSTAR} Data Reduction}

\textit{NuSTAR} \citep{harrison_2013_nustar} telescope consists of two identical Focal Plane Modules, \textit{FPMA} and \textit{FPMB}, to observe the X-ray radiations in the energy range $3-79$ keV. 
The \textit{NuSTAR} archival data are obtained from NASA’s HEASARC archive$^{\ref{web:heasarc}}$, which are reprocessed using \texttt{NuSTARDAS v2.1.1} available in \texttt{HEASOFT v6.29c}, with the latest \textit{NuSTAR} calibration database (\texttt{CALDB v.20230718})\footnote{\url{https://heasarc.gsfc.nasa.gov/docs/heasarc/caldb/caldb_supported_missions.html}} and the clock correction file. We use \texttt{nupipeline v0.4.9} to produce clean event files from raw data.
Source images were extracted from these clean events using \texttt{XSELECT V2.4m}. A circular region of radius 100 arcsec, centred around the source and away from the source is considered for the source and the background region respectively. Lightcurve and spectra are produced for source and background region using \texttt{nuproducts} module for both \textit{FPMA} and \textit{FPMB}. 
The \textit{NuSTAR} spectra are grouped with a minimum of 30 counts per bin. The energy spectra are fitted jointly using \textit{FPMA} and \textit{FPMB}. We did not add any systematic error for our spectral analysis using \textit{NuSTAR} observation.

\subsection{\textit{MAXI} Data Reduction}

The Gas Slit Camera (\textit{GSC}) \citep{mihara_maxi_2011} onboard the Monitor of All sky X-ray Imaging (\textit{MAXI}) mission has been continuously observing GRS 1915+105 since August 11, 2009. We employ \textit{MAXI} mission data from the \textit{MAXI/GSC} website\footnote{\label{web:maxi}\url{http://maxi.riken.jp/top/lc.html}} to examine the source's light curve in the 2$-$20 keV energy range. Hardness Ratio (HR) are defined as the ratio of count rates in 4$-$10 keV to 2$-$4 keV energy band. The \textit{GSC} counts are further converted to Crab units by dividing each count rate by 3.74 counts/s \citep{motta2021}.


\begin{table*}

    \centering
    \caption{\label{tab:log_table}Observation details of the source GRS 1915+105 observed by \textit{RXTE}, \textit{AstroSat} and \textit{NuSTAR} of $\gamma$, $\delta$ and $\phi$ classes during 1996 to 2017. In the table, Obs No., Mission along with ObsID, MJD start and effective exposure time are mentioned. The detected $r_{det}$ count rate for all observations with hardness ratios and fractional variance are also tabulated. The observations without HFQPO are marked as grey. We did not find the HFQPO feature in the \textit{NuSTAR} observations. See text for details.}
    \begin{tabular}{cccccccccc}
   
    \hline
    \hline
     Obs No. & Mission & ObsID   & MJD    &  Effective    & $r_{det}$ & HR1   & HR2  & F$_{var}$  & HFQPO \\
             &       &   &              &  Exposure (s) & (cts/s)   & (B/A)* & (C/A)* & (\%)  \\
    \hline
&&& \multicolumn{2}{c}{$\gamma$ Class}  \\
    \hline
    1 & \textit{RXTE} & 10408-01-03-00 & 50190.578 & 4744       & 10973    & 1.12   & 0.07 & 11.11  & Yes   \\
    2 & \textit{RXTE} & 10408-01-04-00 & 50193.439 & 8688      & 18361    & 1.29   & 0.09 & 5.77 & Yes   \\
    3 & \textit{RXTE} & 10408-01-05-00 & 50202.845 & 9136      & 11546    & 1.16   & 0.08 & 11.50 & Yes   \\
    4 & \textit{RXTE} & 10408-01-06-00 & 50208.584 & 9600     & 12822    & 1.21   & 0.08 & 7.54 & Yes   \\
    5 & \textit{RXTE} & 10408-01-07-00 & 50217.656 & 9664      & 7522     & 1.10   & 0.07 & 11.70 & Yes   \\
     \rowcolor{gray!15}
    6 & \textit{RXTE} & 20402-01-37-02 & 50646.305 & 2680     & 20644    & 1.18   & 0.07 & 11.36 & No    \\
    \rowcolor{gray!15}
    7 & \textit{RXTE} & 20402-01-37-00 & 50646.423 & 8936     & 21060    & 1.19   & 0.06 & 10.78 & No    \\
    8 & \textit{RXTE} & 20402-01-38-00 & 50649.426 & 7480     & 18993    & 1.18   & 0.06 & 11.01 & Yes   \\
    9 & \textit{RXTE} & 20402-01-39-00 & 50654.028 & 6608     & 18415    & 1.24   & 0.07 & 11.96 & Yes   \\
   10 & \textit{RXTE} & 20402-01-39-02 & 50658.511 & 2032     & 18030    & 1.27   & 0.08 & 10.47 & Yes   \\
   \rowcolor{gray!15}
   11 & \textit{RXTE} & 20402-01-40-00 & 50663.789 & 1168     & 21089    & 1.14   & 0.06 & 14.77 & No    \\
   \rowcolor{gray!15}
   12 & \textit{RXTE} & 40703-01-13-00 & 51299.067 & 3307     & 7496    & 1.18   & 0.09 & 12.08 & No    \\
   \rowcolor{gray!15}
   13 & \textit{RXTE} & 40703-01-30-03 & 51432.964 & 2516     & 8264    & 1.16   & 0.09 & 13.94 & No    \\
   \rowcolor{gray!15}
   14 & \textit{RXTE} & 40115-01-07-00 & 51440.608 & 4950     & 13243    & 1.18   & 0.06 & 7.81 & No    \\
   15 & \textit{RXTE} & 93701-01-01-00 & 54285.950 & 1536      & 7680     & 0.72   & 0.03 & 9.38 & Yes    \\   
16 & \textit{NuSTAR} & 30102037004 & 57509.889 & 7509  &  1767 & 1.02 & 0.04 & 25.36  &  $-$    \\  
17 & \textit{AstroSat} & G07\_046T01\_9000001534 (10583) & 58008.080 & 1816 & 7776 &
  0.90 & 0.07 & 10.21  & Yes     \\
18 & \textit{NuSTAR} & 30302020002 & 58018.827 & 7074 & 1959 & 1.00 & 0.04 & 12.33 & $-$                       \\  
  \hline
&&& \multicolumn{2}{c}{$\delta$ Class}  \\
    \hline

    \rowcolor{gray!15}
19 & \textit{RXTE}     & 10408-01-14-00 & 50246.004 & 1888  & 25662 & 1.21 & 0.06 & 20.41 & No       \\

 \rowcolor{gray!15}
20 & \textit{RXTE}     & 10408-01-17-00 & 50256.611 & 3408  & 14365 & 1.08 & 0.06 & 26.31 & No       \\

 \rowcolor{gray!15}
21 & \textit{RXTE}     & 10408-01-17-03 & 50259.080 & 1968  & 15828 & 1.11 & 0.06 & 27.60 & No       \\
    
    \rowcolor{gray!15}
22 & \textit{RXTE}     & 20402-01-41-00 & 50679.238 & 2192  & 24438 & 1.20 & 0.06 & 14.49 & No       \\
    \rowcolor{gray!15}
23 & \textit{RXTE}     & 20402-01-41-01 & 50679.305 & 2624  & 23160 & 1.18 & 0.05 & 14.18 & No       \\
    \rowcolor{gray!15}
24 & \textit{RXTE}     & 20402-01-41-03 & 50679.463 & 1344  & 24836 & 1.19 & 0.06 & 12.17 & No       \\
    \rowcolor{gray!15}
25 & \textit{RXTE}     & 20402-01-42-00 & 50681.796 & 1472  & 24388 & 1.19 & 0.06 & 15.73 & No       \\
26 & \textit{RXTE}     & 20402-01-54-00 & 50763.209 & 9936  & 22356 & 1.20 & 0.06 & 9.51  & Yes      \\
27 & \textit{RXTE}     & 20402-01-55-00 & 50763.229 & 8416  & 16260 & 1.30 & 0.09 & 12.76 & Yes      \\
28 & \textit{RXTE}     & 20402-01-56-00 & 50774.223 & 9056  & 19024 & 1.24 & 0.07 & 9.10  & Yes      \\
29 & \textit{RXTE}     & 20402-01-60-00 & 50804.908 & 11920 & 25188 & 1.22 & 0.06 & 18.04 & Yes      \\
30 & \textit{RXTE}     & 80701-01-28-00 & 52933.627 & 1760  & 9569  & 1.05 & 0.06 & 3.57  & Yes      \\
31 & \textit{RXTE}     & 80701-01-28-01 & 52933.695 & 1520  & 9365  & 1.06 & 0.06 & 3.19  & Yes      \\
32 & \textit{RXTE}     & 80701-01-28-02 & 52933.763 & 1024  & 9042  & 1.07 & 0.07 & 3.37  & Yes      \\
33 & \textit{RXTE}     & 93411-01-01-00 &  54326.708 & 3200  & 3716 & 0.61  &  0.03  & 5.95  & Yes \\
34 & \textit{NuSTAR}   & 10102003002    & 57340.553 & 11919 & 1872   & 0.80  & 0.02 & 7.83  & $-$ \\
35 & \textit{AstroSat} & G05\_189T01\_9000000492 (3819)           & 57551.040  & 2566  & 6834  & 0.91 & 0.07 & 3.09  & Yes      \\
36 & \textit{AstroSat} & G05\_189T01\_9000000492 (3839)           & 57552.350  & 2179  & 7702  & 0.77 & 0.04 & 5.03  & Yes      \\
\rowcolor{gray!15}
37 & \textit{AstroSat} & G05\_189T01\_9000000492 (3845)           & 57552.866 & 3475  & 7802  & 0.76 & 0.03 & 5.29  & No       \\
38 & \textit{NuSTAR}   & 30202033002    & 57553.074 & 22879 & 2080  & 0.81 & 0.02 & 5.70   & $-$ \\
39 & \textit{AstroSat} & G05\_189T01\_9000000492 (3860)           & 57553.880  & 3427  & 6848  & 0.88 & 0.06 & 3.39  & Yes      \\
40 & \textit{AstroSat} & G05\_189T01\_9000000492 (3864)           & 57554.090  & 2363  & 6681  & 0.89 & 0.07 & 3.39  & Yes      \\
41 & \textit{NuSTAR}   & 30202033004    & 57588.595 & 22454 & 537   & 0.51 & 0.01 & 5.54  & $-$ \\
42 & \textit{NuSTAR}   & 30302018002    & 57928.685 & 30615 & 584   & 0.74 & 0.02 & 10.43 & $-$ \\
43 & \textit{NuSTAR}   & 30302020004    & 58038.902 & 21658 & 2207  & 0.72 & 0.01 & 5.61  & $-$ \\

\rowcolor{gray!15}
44 & \textit{AstroSat} & A04\_180T01\_9000001622 (11144)          & 58045.931 & 3218  & 7362  & 0.67 & 0.03 & 5.70   & No       \\
\hline
&&& \multicolumn{2}{c}{$\phi$ Class}  \\
    \hline
    \rowcolor{gray!15}

45 & \textit{RXTE} & 10408-01-09-00 & 50232.530 & 5744 & 8923 & 1.02 & 0.05 & 6.41 & No \\
\rowcolor{gray!15}
46 & \textit{RXTE} & 10408-01-12-00 & 50239.483 & 10600 & 11099 & 1.02 & 0.04 & 8.89  & No \\
\rowcolor{gray!15}
47 & \textit{RXTE} & 10408-01-17-01 & 50256.744 & 3392  & 12329 & 1.05 & 0.05 & 12.11 & No \\

\rowcolor{gray!15}
48 & \textit{RXTE} & 10408-01-19-01 & 50263.550 & 3344  & 5432  & 0.84 & 0.03 & 6.05  & No \\
\rowcolor{gray!15}
49 & \textit{RXTE} & 10408-01-20-01 & 50267.486 & 2936 & 6905 & 0.93 & 0.03 & 7.43 & No   \\
\rowcolor{gray!15}
50 & \textit{RXTE} & 93411-01-01-01 & 54326.316  & 2400 & 2074 & 0.49 & 0.01 & 4.22 & No \\
\hline 
        
    \end{tabular}
    \begin{list}{}{}
         
		\item[*]In the case of \textit{RXTE/PCA}, the energy bands A, B and C are defined as $2-5$ keV, $5-13$ keV and $13-60$ keV respectively. \\
            \item[] For \textit{AstroSat/LAXPC} and \textit{NuSTAR} observation, A, B and C bands are defined as $3-6$ keV, $6-15$ keV and $15-60$ keV respectively.
	\end{list}
    
\end{table*}

\section{Timing Analysis and Results} \label{sec:timing}

We performed a temporal analysis of observations from three variability classes, $\gamma$, $\delta$, and $\phi$ during the ‘canonical’ soft state using data from \textit{RXTE}, \textit{AstroSat}, and \textit{NuSTAR}, spanning from April 17, 1996 to October 19, 2017. This analysis involved generating color-color diagrams (CCDs), power density spectra (PDSs), and time-lag spectra, with the results summarized in Tables~\ref{tab:log_table} and \ref{tab:pds_table}. However, we were unable to detect any significant HFQPO features in the \textit{NuSTAR} observations.

 \subsection{Lightcurve and Colour-Colour Diagram (CCD)} \label{subsec:lightcurve_ccd}
  
We have analysed the \textit{RXTE/PCA} observations by generating background-subtracted 1s binned lightcurves in the 2$-$5 keV, 5$-$13 keV, and 13$-$60 keV energy bands, using all available PCUs. To plot the CCD, we defined soft and hard colours as HR1=B/A and HR2=C/A, respectively, where A, B, and C represent the photon count rates in the 2$-$5 keV, 5$-$13 keV, and 13$-$60 keV energy bands \citep[see][]{belloni2000}.

\begin{figure*}
 \includegraphics[width=0.95\textwidth]{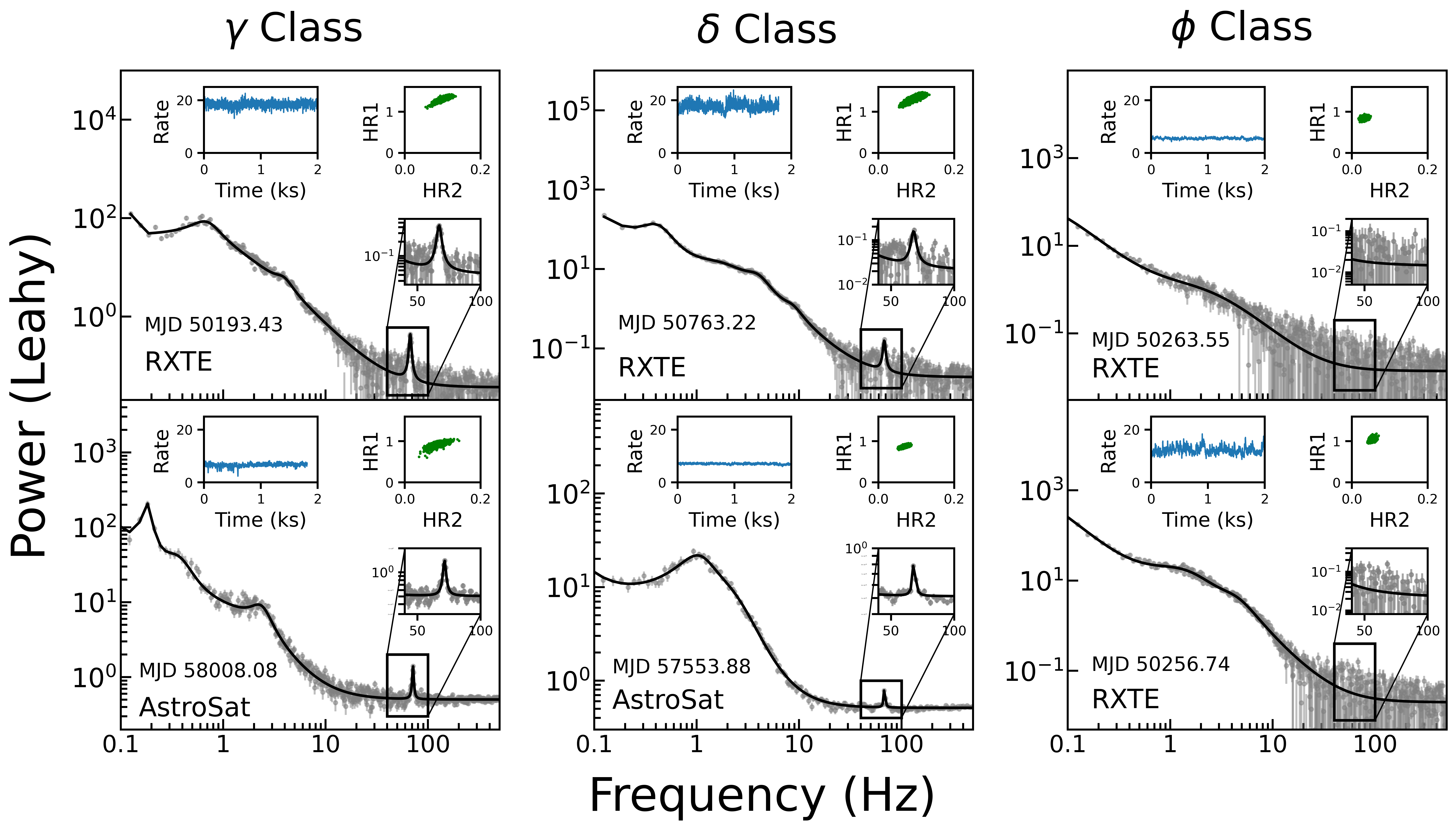}
 \caption{The light curve, CCD, and wide-band PDS of \textit{RXTE} and \textit{AstroSat} observations for the $\gamma$, $\delta$, and $\phi$ classes are shown in three columns (six panels). For each panel, the light curve (top-left inset) is plotted in blue, and the CCD (top-right inset) is plotted in green.
 The wide-band PDS is plotted in grey, with the fitted model in black. A zoomed-in view of the HFQPO feature is shown in the central-right inset of each panel. The MJD of the observation, along with the mission, is shown in the lower-left corner of each panel. See the text for details.
 }
 \label{fig:lc_ccd_pds}
\end{figure*}

We produced 1s binned lightcurves for the \textit{AstroSat} observations, in the 3$-$6 keV (band A), 6$-$15 keV (band B), and 15$-$60 keV (band C) energy bands by combining \textit{LAXPC10} and \textit{LAXPC20} data \citep{sreehari2020,seshadri2022}.  
The background-corrected lightcurve and CCD of $\gamma$, $\delta$ and $\phi$ class observation in the 3$-$60 keV band are shown in the inset of Fig. \ref{fig:lc_ccd_pds}. The 1s binned background subtracted lightcurve for \textit{NuSTAR} observation in 3$-$60 keV energy band is produced by combining \textit{FPMA} and \textit{FPMB}. We produced the background corrected CCD by considering the same energy band as chosen in the case of \textit{AstroSat} observations.

The background-corrected lightcurve, CCD in the 3$-$60 keV energy band for all observations were utilized to calculate the mean value of count rate, soft-colour (HR1), hard-colour (HR2), and fractional variance (F$_{var}$) \citep{vaughan_2003,bhuvana_frac-var2021} of the lightcurve (see Table \ref{tab:log_table}).
For the $\delta$ and $\phi$ class observations, no significant differences are observed in HR1 (mean $\sim$ 1.0), HR2 (mean $\sim$ 0.04), and F$_{var}$ (mean $\sim$ 7\%). In contrast, $\gamma$ class observations show higher average values, with HR1 at 1.12, HR2 at 0.06, and F$_{var}$ at 12\%. 
In $\phi$ class observations, both HR1 and HR2 decreases as the count rate decreases (see Table \ref{tab:log_table}).

\subsection{Power Density Spectrum}
\label{subsection:PDS}

We have used \texttt{GHATS V3.2.0}\footnote{\url{http://astrosat-ssc.iucaa.in/uploads/ghats_home.html}} to generate Power Density Spectrum (PDS) from all the \textit{RXTE/PCA} observations. 
We used the full energy band of 2$-$60 keV in order to detect statistically significant HFQPO features.
We took 32 s Fast Fourier Transform (FFT) segments and averaged over all FFT segments to produce the final PDS in Leahy normalization \citep{leahy1983}.
Dead time corrected Poisson noise was subtracted from the PDS following \cite{zhang1995} and \cite{jahoda_2006_pca}. Finally, we applied logarithmic rebinning to the PDS, increasing the size of each frequency bin by a factor of $exp(1/200)$ compared to the previous one.

The PDS from \textit{AstroSat} observations are produced from the 1 ms binned lightcurve combining \textit{LAXPC10} and \textit{LAXPC20} in the energy band 3$-$60 keV using \texttt{powspec} in \texttt{HEASOFT v6.29c}.
We chose 32768 bins per interval, resulting each segment to be 32.768 s long, which are further averaged to obtain the final PDS. We have used geometric rebin factor of 1.03 for the power spectral analysis. Finally, the dead-time corrected Poisson noise is subtracted following \cite{agrawal2018} and produced the final PDS in Leahy normalization.

HFQPOs are detected for the $\gamma$ (10 observations) and $\delta$ (12 observations) class observations of \textit{RXTE} and \textit{AstroSat}. The $\phi$ variability class observations do not show HFQPO in the PDS. 
We also attempted to extract the cross-power density spectrum (cross-PDS) from \textit{NuSTAR} observations using the Python-based open-source code \texttt{stingray v2.0} \citep{huppenkothen_stingray2019}, as this method minimizes \textit{NuSTAR}'s dead-time effects \citep{bachetti2015}. However, we did not detect any HFQPOs, likely due to the typically low rms amplitude of the HFQPO signal in the cross-PDS. However, the significantly higher spectral resolution of \textit{NuSTAR} data \citep{ratheesh2021}, combined with quasi-simultaneous \textit{AstroSat} observations, enables a more detailed study of the source's spectral features, regardless of the presence or absence of HFQPOs during the canonical soft state.
We fitted the wide-band ($0.1-500$ Hz) PDS of all observations using the model combination of \texttt{Lorentzian} and \texttt{powerlaw} (to model the high frequency Poisson noise) in \texttt{XSPEC V12.12.0}. 
In Fig. \ref{fig:lc_ccd_pds}, we present the model fitted PDS for $\gamma$, $\delta$ and $\phi$ variability classes with a zoomed in view of the HFQPO or without HFQPO feature shown in the inset of each panel.
The best fitted model parameters of the HFQPO feature, including centroid frequency, FWHM, normalization, and estimated parameters such as significance ($\sigma$), HFQPO$_{rms}$ and the Total$_{rms}$ are tabulated in the Table \ref{tab:pds_table}.
The significance of the HFQPO is calculated as the ratio of the \texttt{Lorentzian} normalization of the HFQPO to its negative error \citep{belloni2012}. The percentage rms of the HFQPO in Leahy normalization is calculated by taking the square root of the ratio of the norm of \texttt{Lorentzian} to the mean count rate and then multiplying it by 100 \citep{van_der_klis1988_lag,wang_2024_rms_formula}.
The total percentage rms of the PDS is calculated in the wide frequency range $0.1-500$ Hz \citep[see][and references therein]{geethu_2022}.

We find that the total percentage rms of the PDS for all observations ranges from $5.04-20.87\%$. The centroid frequency of the HFQPO lies in the range 65.07$-$71.38 Hz with a percentage rms of 0.38$-$2.14\%.
It can be noted that during $\phi$ class observations, the Total$_{rms}$\% of the PDS varies linearly with the count rate (see Table \ref{tab:log_table} and \ref{tab:pds_table}).


\begin{table*}
  \centering
    \caption{\label{tab:pds_table}Details of the best fitted PDS parameters of $\gamma$, $\delta$ and $\phi$ class observations of GRS 1915+105 in $2-60$ keV energy range are presented. All PDS are fitted in the wide-band frequency range $0.1-500$ Hz in Leahy space. All the errors are calculated in 68\% confidence range. Time-lags are calculated for $5-25$ keV w.r.t $2-5$ keV and tabulated in units of milliseconds. The observations without HFQPO are marked as grey. The Fourier analysis for \textit{NuSTAR} observations (Obs No. 16, 18, 34, 38, 41, 42, 43) is not performed, hence not tabulated here. See text for details.}
    \resizebox{19cm}{!}{
    \begin{tabular}{ccccccccc}
    \hline
    \hline
    Obs No. & MJD &
    HFQPO$_{freq}$ (Hz) & FWHM (Hz) & HFQPO$_{norm}$  & Significance ($\sigma$) & HFQPO$_{rms}$\%  & Total$_{rms}$\%  & Time-lag (ms)    \\
    \hline
&&& \multicolumn{3}{c}{$\gamma$ Class}  \\
    \hline
1 & 50190.578$^{a}$ & 69.38 $_{-0.69}^{+0.80}$ & 4.21 $_{-2.13}^{+2.80}$ & 0.44 $_{-0.15}^{+0.17}$ & 2.93 & 0.64 $\pm$ 0.12  & 10.61 $\pm$ 0.08 & $-$ \\

2 & 50193.439$^{a}$ & 67.32 $_{-0.13}^{+0.17}$ & 3.25 $_{-0.39}^{+0.36}$ & 1.21 $_{-0.09}^{+0.10}$ & 13.44 &  0.81 $\pm$ 0.03 & 9.07 $\pm$ 0.16 & $-$\\

3 & 50202.845 & 66.03 $_{-0.30}^{+0.32}$ & 2.82 $_{-0.47}^{+0.62}$ & 0.52 $\pm$ 0.08 & 6.50 & 0.67 $\pm$ 0.05 & 10.91 $\pm$ 0.05 & 2.84 $\pm$ 0.92  \\

4 & 50208.584 & 65.23 $_{-0.09}^{+0.15}$ & 3.63 $_{-0.28}^{+0.27}$ & 1.75 $\pm$ 0.10 & 17.50 & 1.16 $\pm$ 0.03 & 10.29 $\pm$ 0.04 & 1.59 $\pm$ 0.34  \\

5 & 50217.656 & 66.64 $_{-0.32}^{+0.31}$ & 2.13 $_{-0.72}^{+1.02}$ & 0.26 $_{-0.06}^{+0.07}$ & 4.33 &  0.59 $\pm$ 0.07  & 10.36 $\pm$ 0.06 & 6.44 $\pm$ 1.72  \\

\rowcolor{gray!15}
6 & 50646.305 &   $-$    &  $-$   &  $-$  & $-$  & $-$ & 11.71 $\pm$ 0.06 &    $-$   \\

\rowcolor{gray!15}
7 & 50646.423 &   $-$    &  $-$   &  $-$  &  $-$ & $-$ & 11.26 $\pm$ 0.06  &   $-$  \\


8 & 50649.426 & 67.21 $_{-0.53}^{+0.53}$ & 2.80 $_{-1.16}^{+1.60}$ & 0.28 $_{-0.09}^{+0.10}$ & 3.11 & 0.38 $\pm$ 0.06 & 10.69 $\pm$ 0.07 &  6.81 $\pm$ 0.33 \\

9 & 50654.028 & 67.84 $_{-0.71}^{+0.76}$ & 7.40 $_{-2.64}^{+2.60}$ & 1.32 $\pm$ 0.29 & 4.55 & 0.84 $\pm$ 0.09 & 10.81 $\pm$ 2.22 & 4.96 $\pm$ 0.61  \\

10 & 50658.511 & 69.06 $\pm$ 0.62 & 5.00 $_{-1.55}^{+2.08}$ & 1.01 $_{-0.24}^{+0.27}$ & 4.21 & 0.74 $\pm$ 0.09 & 9.55 $\pm$ 0.19 & 5.80 $\pm$ 1.30  \\

\rowcolor{gray!15}
11 & 50663.789 &  $-$  & $-$  &  $-$  & $-$  &  $-$ & 15.93 $\pm$ 0.36  & $-$        \\

\rowcolor{gray!15}
12 & 51299.067 &  $-$  & $-$  & $-$  & $-$  &  $-$  & 12.91 $\pm$ 0.12   &  $-$ \\

\rowcolor{gray!15}
13 & 51432.964 &  $-$  & $-$  & $-$  & $-$  & $-$ & 15.08 $\pm$ 0.17  & $-$ \\

\rowcolor{gray!15}
14 & 51440.608$^{\dagger}$ &  $-$  & $-$ & $-$  & $-$  & $-$ & 7.43 $\pm$ 0.44 & $-$  \\

15 & 54285.950 & 71.04 $_{-0.72}^{+0.67}$ & 3.38 $_{-1.88}^{+1.90}$ & 0.69 $_{-0.26}^{+0.25}$ & 2.65 & 0.95 $\pm$ 0.18 & 10.37 $\pm$ 0.12 & 6.82 $\pm$ 1.05  \\

17 & 58008.080 & 71.38 $_{-0.08}^{+0.13}$ & 2.50 $_{-0.28}^{+0.29}$ & 3.58 $\pm$ 0.27 & 13.26 & 2.14 $\pm$ 0.08  & 20.19 $\pm$ 0.06  & -1.68 $\pm$ 0.20 \\

\hline

&&& \multicolumn{3}{c}{$\delta$ Class}  \\
    \hline

 \rowcolor{gray!15}
19 & 50246.004 & $-$  & $-$   &  $-$ &  $-$  &  $-$  & 9.50 $\pm$ 2.10 & $-$ \\   

 \rowcolor{gray!15}
20 & 50256.611 & $-$  & $-$   &  $-$ &  $-$  &  $-$  & 12.19 $\pm$ 2.04 & $-$ \\ 

 \rowcolor{gray!15}
21 & 50259.080 & $-$  & $-$   &  $-$ &  $-$  &  $-$  & 11.34 $\pm$ 0.20 & $-$ \\ 

\rowcolor{gray!15}
22 & 50679.238 & $-$  & $-$   &  $-$ &  $-$  &  $-$  & 5.41 $\pm$ 0.06 & $-$ \\

\rowcolor{gray!15}
23 & 50679.305 & $-$  & $-$   &  $-$ &  $-$  & $-$  &  5.34 $\pm$ 0.06 & $-$ \\

\rowcolor{gray!15}
24 & 50679.463 & $-$  & $-$  &  $-$  &  $-$  & $-$  & 5.18 $\pm$ 0.10 & $-$ \\

\rowcolor{gray!15}
25 & 50681.796 & $-$  &  $-$ &  $-$ &   $-$  & $-$  & 5.27  $\pm$ 0.73 & $-$ \\

26 & 50763.209 & 66.00 $_{-0.55}^{+0.46}$ & 3.85$_{-1.59}^{+2.37}$ & 0.37 $_{-0.10}^{+0.12}$ & 3.70  & 0.40 $\pm$ 0.05  & 6.07 $\pm$ 0.02 & 7.16 $\pm$ 0.34 \\

27 & 50763.229 & 67.98$_{-0.24}^{+0.28}$ & 3.04$_{-0.57}^{+0.64}$ & 0.61$_{-0.08}^{+0.09}$ & 7.63  & 0.61 $\pm$ 0.04 & 9.24 $\pm$ 0.94 & 5.02 $\pm$ 0.49 \\

28  & 50774.223 & 68.69 $\pm$ 0.29 & 3.15$_{-0.84}^{+1.06}$ & 0.49$_{-0.09}^{+0.10}$ & 5.44 & 0.50 $\pm$ 0.04 & 7.23 $\pm$ 0.03 & 5.99 $\pm$ 0.78 \\

29  & 50804.908 & 65.07$_{-0.32}^{+0.27}$ & 3.07$_{-0.76}^{+0.86}$  & 0.42$\pm$ 0.07 & 6.00  & 0.41 $\pm$ 0.03 & 5.04 $\pm$ 0.22 & 7.55$\pm$ 0.36 \\
 
30  & 52933.627 & 68.29 $\pm$ 0.09 & 2.37$_{-0.17}^{+0.25}$ & 3.19$_{-0.15}^{+0.21}$ & 21.27  & 1.82 $\pm$ 0.04 & 7.64 $\pm$ 0.09 & 1.80 $\pm$ 0.19 \\

31  & 52933.695 & 67.90 $\pm$ 0.08 & 2.44$_{-0.18}^{+0.25}$  & 3.84$_{-0.20}^{+0.23}$ & 19.20  & 2.03 $\pm$ 0.05 & 7.32 $\pm$ 0.08 & 1.74 $\pm$ 0.24 \\

32  & 52933.763 & 67.96$_{-0.15}^{+0.12}$  & 2.45$_{-0.24}^{+0.36}$ & 3.69$_{-0.23}^{+0.31}$ & 16.04 & 2.02 $\pm$ 0.06 &  6.83 $\pm$ 0.07 & 1.84 $\pm$ 0.24 \\

33  & 54326.708 & 66.32$_{-0.77}^{+0.72}$  &  5.39$_{-1.74}^{+2.50}$  &  0.75$_{-0.19}^{+0.22}$ & 3.95  &  1.42 $\pm$ 0.18 & 7.57 $\pm$ 0.31 & 1.07 $\pm$ 2.07 \\

35  & 57551.040  & 68.02$_{-0.05}^{+0.06}$ & 1.49 $\pm$ 0.45 & 1.33 $\pm$ 0.16 & 8.31  & 1.39 $\pm$ 0.08  & 20.7 $\pm$ 0.03 & -1.06 $\pm$ 0.17\\

36  & 57552.350  & 68.50$_{-0.29}^{+0.34}$ & 2.00*  & 0.50$_{-0.11}^{+0.12}$  & 4.55  & 0.81 $\pm$ 0.09 & 19.76 $\pm$ 0.03 & -0.49 $\pm$ 0.16 \\

\rowcolor{gray!15}
37 & 57552.866 & $-$ & $-$ & $-$ & $-$ & $-$ & 20.14 $\pm$ 0.03 & $-$   \\
  
39  & 57553.880  & 68.08$_{-0.24}^{+0.18}$ & 2.55$_{-0.65}^{+0.70}$ & 1.31$_{-0.17}^{+0.18}$ & 7.71  & 1.38 $\pm$ 0.09 & 20.87 $\pm$ 0.03 & -0.74 $\pm$ 0.12 \\
  
40  & 57554.090  & 68.13$_{-0.11}^{+0.12}$ & 2.00*  & 1.35 $\pm$ 0.12  & 11.25 & 1.42 $\pm$ 0.06  & 20.73 $\pm$ 0.03 & -1.00 $\pm$ 0.16 \\

\rowcolor{gray!15}
44  & 58045.931 &  $-$     &   $-$    &  $-$    &   $-$   & $-$       & 19.71 $\pm$ 0.03  & $-$   \\

\hline

&&& \multicolumn{3}{c}{$\phi$ Class}  \\
    \hline

\rowcolor{gray!15}
45 & 50232.530 & $-$ & $-$ & $-$ & $-$ & $-$ & 7.18 $\pm$ 0.21 & $-$ \\
 
\rowcolor{gray!15}
46 & 50239.483 & $-$ & $-$ &  $-$  &  $-$  &  $-$  & 7.74 $\pm$ 0.03 & $-$  \\

\rowcolor{gray!15}
47 & 50256.744 & $-$ & $-$ &  $-$  &  $-$  &  $-$  & 8.79 $\pm$ 0.06 &  $-$ \\

\rowcolor{gray!15}
48 & 50263.831 & $-$ & $-$ &  $-$  &  $-$  &  $-$  & 5.55 $\pm$ 0.15 & $-$ \\


\rowcolor{gray!15}
49 & 50267.486 & $-$ & $-$ & $-$ & $-$ & $-$ & 6.92 $\pm$ 0.68 & $-$   \\

\rowcolor{gray!15}
50 & 54326.316 & $-$ & $-$ & $-$ & $-$ & $-$ & 5.36 $\pm$ 0.33 & $-$   \\
\hline

\end{tabular}
 }

\begin{list}{}{}
         \item[$a$] Time-lag could not be calculated due to the unavailability of channel ranges. 

         \item[$\dagger$] Total rms is calculated in the energy range 13$-$60 keV due to unavailibility of high resolution data in 2$-$13 keV range.
         
	\item[*] Fixed parameter

	\end{list}
\end{table*}

\subsection{Time-lag Spectrum} \label{subsec:time-lag}

\begin{figure}
	\includegraphics[width=\columnwidth]{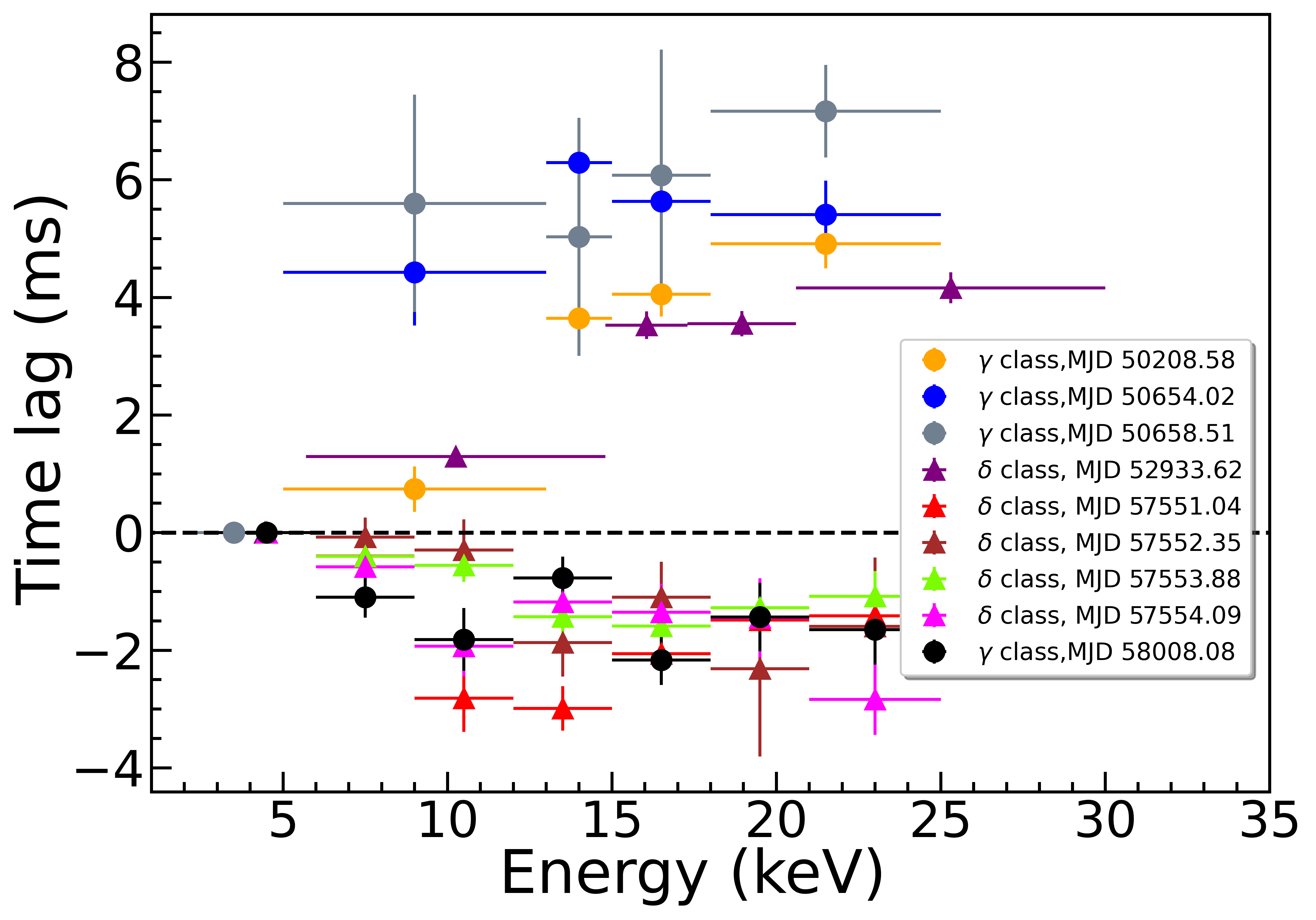}
    \caption{The energy-dependent time-lags at HFQPOs are plotted. A hard-lag feature is observed for all the \textit{RXTE} observations, whereas the \textit{AstroSat} observations show a soft-lag associated with the HFQPO. The variability class, along with the MJD, is shown in the legend. See the text for details. }
    \label{fig:ene_dep_time-lag}
\end{figure}

We computed the cross-spectrum, defined as $C(j)=X^{*}_{1}(j)X_{2}(j)$, where $X_{1}(j)$ and $X_{2}(j)$ are the complex Fourier coefficients corresponding to two energy bands at a frequency $\nu (j)$. The term $X^{*}_{1}(j)$ refers to the complex conjugate of $X_{1}(j)$ \citep{van_der_klis1987}.
The time-lag is calculated from the argument of the complex cross-spectra \citep[see][for more details]{vaughan_nowak1997} for different energy bands. We produced the cross-spectra using \texttt{GHATS V3.2.0} of 5$-$25 keV energy band with respect to 2$-$5 keV band with FFT length of 32 s and averaged the time-lag over the FWHM of HFQPO following \cite{reig2000} (see the Table \ref{tab:pds_table}). 
The time delay, or lag, in the arrival of hard photons compared to soft photons is termed a hard-lag with the reverse known as a soft-lag \citep{van_der_klis1988_lag}.
We have calculated the time-lag for each of 
four different energy bands, 5$-$13 keV, 13$-$15 keV, 15$-$18 keV and 18$-$25 keV with respect to 2$-$5 keV band. The time-lag for $\gamma$ and $\delta$ class observations from \textit{RXTE/PCA} is plotted as a function of energy and shown in Fig. \ref{fig:ene_dep_time-lag}. 
The time-lag analysis for \textit{AstroSat/LAXPC} observations was performed following \cite{prajjwal_2024} using \texttt{LAXPCsoftware} and combining data from both \textit{LAXPC10} and \textit{LAXPC20}. The time-lag of 6$-$25 keV energy band is calculated with respect to 3$-$6 keV band for all \textit{AstroSat} observations and tabulated in Table \ref{tab:pds_table}. 
Energy dependent time-lag spectra for the \textit{AstroSat} observation of the $\gamma$ and $\delta$ classes were generated for the 6$-$9 keV, 9$-$12 keV, 12$-$15 keV, 15$-$18 keV, 18$-$21 keV and 21$-$25 keV bands, relative to the softest energy band (3$-$6 keV) and are shown in Fig. \ref{fig:ene_dep_time-lag}.

For \textit{RXTE} observations (MJD 50190$-$54326), we found a hard-lag in the range of $1.59-7.55$ ms for $5-25$ keV photons w.r.t the $2-5$ keV photons. In contrast, for \textit{AstroSat} observations (MJD 57551$-$58008), a soft-lag ranging from $0.49-1.68$ ms was observed for $6-25$ keV photons w.r.t $3-6$ keV photons (see Table \ref{tab:pds_table}).
Fig. \ref{fig:ene_dep_time-lag} shows that the hard-lag for \textit{RXTE} observations of both $\gamma$ and $\delta$ classes increase with energy, consistent with the findings of \cite{cui1999} and \cite{mendez2013} (Obs 4 and combining Obs 30, 31 and 32 in Table \ref{tab:log_table}). However, in case of \textit{AstroSat} observations, a soft-lag is observed to increase with energy as first reported by \cite{prajjwal_2024}.

\section{Spectral Analysis and results}
\label{subsec:spectral_analysis}
We have analysed wide-band energy spectra using \texttt{XSPEC V12.12.0} \citep{k.arnaud_1996_xspec} by combining different detectors of an instrument (\textit{RXTE/PCA-HEXTE, AstroSat/SXT-LAXPC} and \textit{NuSTAR/FPMA-FPMB}). We used a cross-normalization constant, which is available as \texttt{constant} in \texttt{XSPEC}. 
We started with \textit{NuSTAR} observations for wide-band (3$-$50 keV) spectral study, when available, due to its superior spectral resolution compared to \textit{RXTE} and \textit{AstroSat}.

We initially adopted the model, \texttt{constant$\times$Tbabs$\times$gabs$\times$ (nthComp+powerlaw)} (hereafter, model M1) to fit the wide-band energy spectrum of observation 42 (MJD 57928.685). The \texttt{Tbabs} \citep{wilms2000} accounts for the galactic absorption, while \texttt{nthComp} \citep{zdziarski1996} calculates the thermal Comptonization of the medium. The hydrogen column density (nH) was kept fixed at $6\times10^{22}$ atoms/cm$^{2}$ \citep{muno1999,sreehari2020}.
We used the Gaussian absorption component \texttt{gabs} \citep{nakajima_gabs_2010} to model an absorption feature at $\sim$ 7 keV \citep{athulya2023}.

\begin{figure}
\includegraphics[width=\columnwidth]{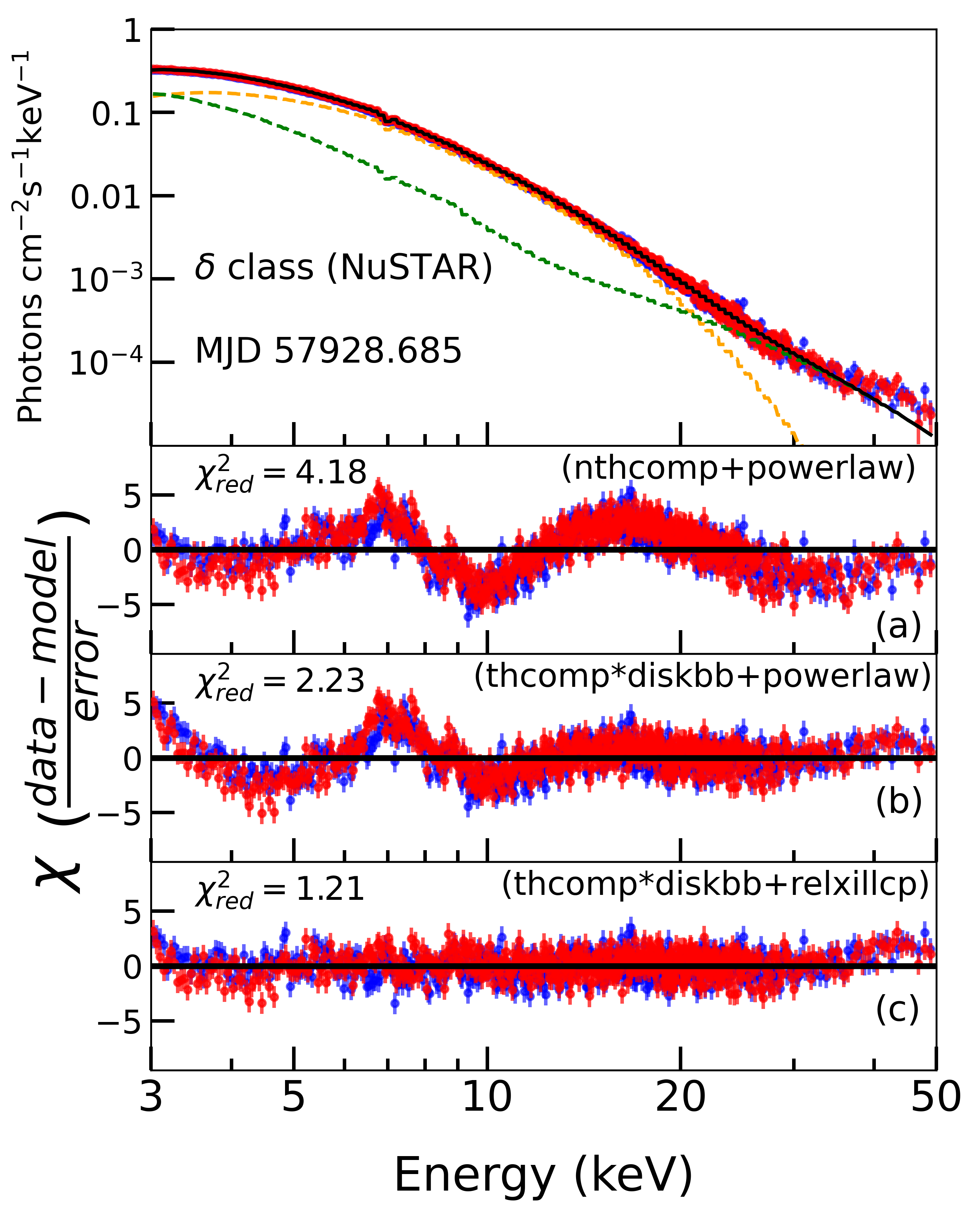}
\caption{The wide-band unfolded energy spectrum of the source GRS 1915+105, using \textit{NuSTAR} observation during the $\delta$ class, is fitted with model (M3), \texttt{constant$\times$Tbabs$\times$gabs$\times$ (thcomp*diskbb+relxillCp)}. The spectra are jointly fitted using \textit{FPMA} (red) and \textit{FPMB} (blue), and the residuals for three different model combinations are shown, along with the obtained $\chi_{red}^{2}$. See the text for details.}
\label{fig:spectrum}
\end{figure}

Initially, the seed photon temperature ($kT_{bb}$) was fixed at 0.1 keV, resulting in a poor reduced $\chi^{2}$ ($\chi^{2}_{red}=\chi^{2}/dof$) of $5149.02/1229=4.18$. The residual, shown in panel (a) in Fig. \ref{fig:spectrum}, reveals a clear broad iron line feature at $\sim$ 7 keV, and a broad hump at $\sim$ 20 keV, indicative of the reflection component. Allowing $kT_{bb}$ to vary improved the fit to $\chi^{2}_{red} = 3.19$ with $kT_{bb}$ increasing to 1.36 keV, although the residual features remained similar. We then adopted the model \texttt{constant$\times$Tbabs$\times$gabs$\times$(thcomp*diskbb+powerlaw)} (hereafter, model M2), replacing the \texttt{nthComp} component with the \texttt{thcomp} convolution model for thermal Comptonization \citep{zdziarski_2020_thcomp}. This resulted in an improved reduced $\chi^2$ of 2.23, though the residuals still showed the presence of the reflection feature (see panel (b) in Fig. \ref{fig:spectrum}).
We incorporated the relativistic blurred iron line feature in model M2 using \texttt{relline} \citep{dauser_relline_2010} from the \texttt{RELXILL V2.3}\footnote{\label{web:relxill}\url{https://www.sternwarte.uni-erlangen.de/~dauser/research/relxill/}} \citep{dauser2022}, achieving a $\chi^{2}_{red}$ of 1.41. However, the residual still showed a Compton hump, leading us to replace the \texttt{powerlaw} component with the relativistic reflection model \texttt{relxillCp}.
The final adopted model  \texttt{constant$\times$Tbabs$\times$gabs$\times$(thcomp*diskbb+relxillCp)} (hereafter M3), incorporates the self consistent reflection model \texttt{relxillCp} to account for both thermal Compton and reflected emissions from disc. The reflection fraction was fixed at $-1$ to segregate the disc's reflected emission, with thermal Comptonization parameters tied to \texttt{thcomp} \citep{sridhar_2020_relxill, sajad_2024_grs_relxill}. The spin parameter was fixed to 0.998 \citep{sreehari2020}.
We assumed equal emissivity indices ($q_{1}=q_{2}=3$), fixed the break radius at 15$R_{g}$ ($R_{g}=GM/c^{2}$) \citep[see][]{bhuvana2023_relxill}, and the outer disc radius at 400 $R_{g}$, allowing $R_{in}$ (in $R_{ISCO}$) and other parameters to vary freely. It can be noted that the \texttt{gabs} component was not used for \textit{RXTE} and \textit{AstroSat} spectra due to the absence of absorption feature at $\sim$ 7 keV. Moreover, we used \texttt{smedge} \citep{ebisawa_smedge1994} component as per requirement.

The best fit yielded $\chi^{2}_{red} = 1.21$, with the residuals shown in panel (c) of Fig. \ref{fig:spectrum}. All observations were fitted using model M3. Error estimation was performed through Markov Chain Monte Carlo (MCMC) simulations with the Goodman-Weare algorithm \citep{goodman_weare_MCMC_2010} in \texttt{XSPEC}, using 50 walkers, a chain length of 20,000, and a burn length of 10,000. Best-fit parameters with 90\% confidence errors are listed in Table \ref{tab:spectral_parameter}.

In the soft state, the inner disc temperature is observed to be higher, ranging from  1.07 to 1.90 keV, the electron temperature ($kT_{e}$) spans 2.03$-$3.61 keV, and the photon index ($\Gamma$) varies between 1.58 and 4.45. The covering fraction ($f_{cov}$) ranges from 0.08 to 0.91. The inner disc radius ($R_{in}$) is found to be located near the innermost stable circular orbit (1.78 to 13.70 $R_{ISCO}$). A high ionization ($log\xi$ = 1.28$-$4.7) and higher plasma density ($logN$ = 15$-$20) are observed in the disc.
We find that the coronal temperature ($kT_{e}$) is low, as expected for the soft-state observations of GRS 1915+105. Our findings align with previous studies on the source in this state \citep[][]{ueda_2009,pahari_pal_grs2010}. We also determined the necessity of the coronal component (\texttt{thcomp}) and found that both \texttt{thcomp} and \texttt{relxillCp} are essential for achieving the best fit and accurately modelling the spectra. Furthermore, freezing $kT_e$ at a higher value ($\sim$ 100 keV) resulted in an unusually low covering fraction ($ \sim 10^{-12}$), suggesting that the Comptonization process becomes insignificant. Thus, despite the low coronal temperature, our analysis confirms that Comptonization remains a crucial component in explaining the observed energy spectra.
The source's inclination angle ranges between 61.19$^{\circ}$ and 74.52$^{\circ}$, with an average of 65.95$^{\circ}$, consistent with previous findings of $\sim$ 65$^{\circ}$ \citep{zdziarski2014}.
The unabsorbed bolometric flux of GRS 1915+105 during the $\gamma$, $\delta$, and $\phi$ classes ranges from 2.28 to 13.32 $\times 10^{-8}$ erg cm$^{-2}$ s$^{-1}$. Assuming a mass of 12.4 M$_{\odot}$ and a distance of 8.6 kpc \citep{reid2014}, the bolometric luminosity is 12.86\% to 75.16\% of the Eddington limit.
The accretion disc contributes 56.65\% to the flux in the 3-50 keV range, indicating a thermally dominated soft state of the source. The contributions from Comptonization and reflection are 28.50\% and 14.83\% of the total flux, respectively.
We calculated the optical depth ($\tau$) to understand the nature and efficiency of the Comptonizing medium near the source. In the `canonical' soft state, the optical depth ranges from 3.21 to 17.35. All estimated parameters are listed in Table \ref{tab:spectral_parameter}.

\section{Spectro-Temporal Correlation}
\label{sec:spectro-temporal_corr}


The in-depth analysis of the source during the canonical soft state reveals variations in spectral and temporal parameters depending on the presence or absence of HFQPOs in the PDS, based on \textit{RXTE}, \textit{AstroSat}, and \textit{NuSTAR} observations, as summarized in Tables~\ref{tab:pds_table} and \ref{tab:spectral_parameter}. A correlated variability study of these spectro-temporal parameters could shed light on the mechanism responsible for HFQPO generation.

\subsection{Spectro-Temporal Parameters and HFQPO}
\label{subsec:spectro-temp-para}

The HFQPOs were detected, when the photon index $\Gamma$ exhibits higher values, lying in the range $2.30-4.45$. The electron temperature ($kT_{e}$) varies from $2.31-3.61$ keV, with a mean of 2.93 keV. The inner disc radius ($R_{in}$) was comparatively larger, ranging from 2.72 to 7.70 $R_{ISCO}$, with an average of 4.65 $R_{ISCO}$. Notable variations were observed in the covering fraction ($0.52 \leq f_{cov} \leq 0.91$) and optical depth ($3.12 \leq \tau \leq 8.98$) of the Comptonizing medium.  
The average disc, Compton, and reflection fluxes are found to be 50.88\%, 37.39\%, and 11.71\% of the total flux, respectively.

In contrast, during observations without HFQPO, we find  $\Gamma$ that exhibits a lower value, ranging from 1.58 to 2.39.
The electron temperature ($kT_{e}$) lies between $2.03-2.97$ keV with a mean of 2.37 keV indicating a relatively cooler corona and inner disc radius ($R_{in}$) varies from 1.78 to 4.58 $R_{ISCO}$, with an average of 2.66 $R_{ISCO}$. 
We also find for observations without HFQPOs, the $f_{cov}$ and $\tau$ are found to be in the range between $0.08 \leq f_{cov} \leq 0.33$ and $9.03 \leq \tau \leq 17.35$, respectively. The average contributions of the disc, Compton, and reflection fluxes to the total flux are calculated as 64.92\%, 19.47\%, and 15.60\%, respectively. However, no significant variation is observed in the inner disc temperature ($T_{in}$), regardless of whether HFQPOs are present or absent.

The correlation between covering fraction ($f_{cov}$) and optical depth ($\tau$) is shown in Fig. \ref{fig:spec_correlation0} according to the Table \ref{tab:pds_table} and \ref{tab:spectral_parameter}. The colour of each data point represents the inner disc radius of the accretion disc in the unit of $R_{ISCO}$. The pink region represents the region where HFQPOs were detected and the grey region indicates the region where HFQPO was absent. 
Although the HFQPO feature was not detected in 5 \textit{NuSTAR} observations due to its high dead-time effect, Fig. \ref{fig:spec_correlation0} shows that 3 of the observations lie within the HFQPO region (pink), while the remaining 2 are in the no-HFQPO region (grey).


\begin{landscape}



\begin{table}
\centering
\caption{\label{tab:spectral_parameter}Model parameters of the observations fitted with model \texttt{constant$\times$Tbabs$\times$(thcomp*diskbb+relxillCp)}. All errors are calculated with 90\% confidence level. The observations without HFQPO are highlighted in grey. See text for details.}
\resizebox{24cm}{!}{
\begin{tabular}{ccccccccccccccccc}
\hline
\hline
Obs No. & MJD    &  $kT_{e}$  &  $\Gamma_{th}$ & $f_{cov}$  & $T_{in}$ & $R_{in}$ & $log\xi$ & $logN$ & relxill$_{norm}$ & $\chi^{2}$/dof & F$_{Compton}^{\dagger}$ & F$_{Disc}^{\dagger}$ & F$_{Refl}^{\dagger}$ & F$_{Bol}^{\ddagger}$  & Optical Depth ($\tau$)     \\
      &  &  (keV)  &    &  & (keV)  & ($R_{ISCO}$)  &   & (cm$^{-3}$) \\

	 \hline
&&&&&&& \multicolumn{3}{c}{$\gamma$ Class}  \\
      \hline

1 & 50190.578 & 2.73$_{-0.11}^{+0.13}$ & 2.35$_{-0.11}^{+0.17}$ & 0.52$_{-0.06}^{+0.11}$ & 1.38$_{-0.05}^{+0.02}$ & 3.38$_{-1.96}^{+8.54}$ & 2.29$_{-1.75}^{+0.60}$ & 17.05$_{-1.79}^{+1.93}$ & 0.19$_{-0.07}^{+0.14}$ & 54.80/78 & 1.78 $\pm$ 0.01 & 2.59 $\pm$ 0.01 & 0.12 $\pm$ 0.01 & 7.56 $\pm$ 0.02 & 8.39 $\pm$ 0.55 & \\ 

2 & 50193.439 & 2.94$_{-0.07}^{+0.20}$ & 2.56$_{-0.05}^{+0.17}$ & 0.67$_{-0.02}^{+0.06}$ & 1.53$_{-0.02}^{+0.03}$ & 3.75$_{-1.64}^{+2.37}$ & 3.92$_{-0.11}^{+0.20}$ & 17.08$_{-0.21}^{+0.19}$ & 0.18$_{-0.03}^{+0.07}$ & 73.29/75 & 1.59 $\pm$ 0.02 & 2.48 $\pm$ 0.02 & 0.57 $\pm$ 0.01 & 8.52 $\pm$ 0.02 & 7.16 $\pm$ 0.20 & \\ 

3 & 50202.845 & 3.02$_{-0.06}^{+0.05}$ & 2.58$_{-0.04}^{+0.02}$ & 0.91$_{-0.08}^{+0.07}$ & 1.62$_{-0.12}^{+0.07}$ & 4.17$_{-1.27}^{+0.56}$ & 4.15 $\pm$ 0.07 & 17.03$_{-0.06}^{+0.03}$ & 0.23$_{-0.03}^{+0.01}$ & 69.59/76 & 1.00 $\pm$ 0.01 & 1.25 $\pm$ 0.01 & 0.81 $\pm$ 0.01 & 6.55 $\pm$ 0.01 & 7.00 $\pm$ 0.12 & \\ 

4 & 50208.584 & 2.74$_{-0.05}^{+0.14}$ & 2.38 $\pm$ 0.01 & 0.52$_{-0.09}^{+0.02}$ & 1.52$_{-0.02}^{+0.07}$ & 4.17$_{-1.68}^{+10.15}$ & 3.83$_{-0.21}^{+0.16}$ & 19.02 $\pm$ 0.08 & 0.039$_{-0.009}^{+0.007}$ & 79.93/74 & 1.21 $\pm$ 0.02 & 2.11 $\pm$ 0.02 & 0.72 $\pm$ 0.01 & 6.84 $\pm$ 0.04 & 8.24 $\pm$ 0.09 & \\ 

5 & 50217.656$^{a}$ & 2.86 $\pm$ 0.11 & 2.72$_{-0.06}^{+0.07}$ & 0.66$_{-0.05}^{+0.07}$ & 1.67 $\pm$ 0.05 & 2.72$_{-0.83}^{+1.79}$ & 1.75$_{-0.06}^{+0.08}$ & 19.00 $\pm$ 0.02 & 0.320$_{-0.05}^{+0.03}$ & 53.81/47 & 0.89 $\pm$ 0.02 & 1.85 $\pm$ 0.02 & 0.41 $\pm$ 0.01 & 9.36 $\pm$ 0.09 & 6.76 $\pm$ 0.24 & \\ 

\rowcolor{gray!15}
6 & 50646.305 & 2.42$_{-0.08}^{+0.10}$ & 2.00$_{-0.03}^{+0.04}$ & 0.19$_{-0.03}^{+0.04}$ & 1.83 $\pm$ 0.04 & 1.78$_{-0.24}^{+0.36}$ & 3.30$_{-0.11}^{+0.12}$ & 18.98$_{-0.04}^{+0.02}$ & 0.13 $\pm$ 0.01 & 66.54/76 & 0.69 $\pm$ 0.04 & 3.20 $\pm$ 0.04 & 1.43 $\pm$ 0.03 & 10.06 $\pm$ 0.04 & 11.17 $\pm$ 0.37 & \\ 

\rowcolor{gray!15}
7 & 50646.423 & 2.46$_{-0.07}^{+0.03}$ & 1.85$_{-0.29}^{+0.08}$ & 0.16 $\pm$ 0.02 & 1.49$_{-0.01}^{+0.04}$ & 2.25$_{-0.96}^{+1.28}$ & 2.30$_{-0.34}^{+0.21}$ & 15.96$_{p}^{+1.78}$ & 0.09$_{-0.01}^{+0.02}$ & 76.72/78 & 1.17 $\pm$ 0.01 & 3.77 $\pm$ 0.01 & 0.23 $\pm$ 0.01 & 8.76 $\pm$ 0.01 & 12.38 $\pm$ 2.86 & \\ 


8 & 50649.426 & 3.61$_{-0.32}^{+0.15}$ & 3.60$_{-0.22}^{+0.07}$ & 0.90$_{-0.12}^{+0.08}$ & 1.40$_{-0.03}^{+0.02}$ & 7.00* & 3.00$_{-0.87}^{+0.07}$ & 15.23$_{-0.01}^{+1.28}$ & 7.70$_{-2.78}^{+3.16}$ & 70.72/69 & 1.52 $\pm$ 0.02 & 3.05 $\pm$ 0.02 & 0.15 $\pm$ 0.01 & 10.76 $\pm$ 0.01 & 4.10 $\pm$ 0.15 & \\ 

9 & 50654.028 & 2.94$_{-0.07}^{+0.11}$ & 2.76$_{-0.07}^{+0.11}$ & 0.76$_{-0.05}^{+0.07}$ & 1.34 $\pm$ 0.02 & 4.52$_{-2.73}^{+2.96}$ & 1.85$_{-0.31}^{+0.28}$ & 18.06$_{-0.19}^{+0.11}$ & 0.52$_{-0.15}^{+0.25}$ & 63.84/78 & 1.90 $\pm$ 0.03 & 2.51 $\pm$ 0.03 & 0.16 $\pm$ 0.01 & 9.35 $\pm$ 0.07 & 6.53 $\pm$ 0.36 & \\ 

10 & 50658.511 & 2.67$_{-0.11}^{+0.14}$ & 2.33$_{-0.05}^{+0.09}$ & 0.54$_{-0.05}^{+0.06}$ & 1.58 $\pm$ 0.06 & 3.94$_{-1.75}^{+3.68}$ & 3.80$_{-0.16}^{+0.22}$ & 17.93$_{-0.17}^{+0.22}$ & 0.10 $\pm$ 0.01 & 96.46/75 & 1.42 $\pm$ 0.02 & 2.45 $\pm$ 0.02 & 0.72 $\pm$ 0.02 & 8.27 $\pm$ 0.04 & 8.60 $\pm$ 0.51 & \\ 

\rowcolor{gray!15}
11 & 50663.789 & 2.48$_{-0.12}^{+0.12}$ & 2.01$_{-0.10}^{+0.04}$ & 0.19$_{-0.03}^{+0.02}$ & 1.69$_{-0.05}^{+0.03}$ & 3.75$_{-1.44}^{+3.70}$ & 3.08$_{-0.08}^{+0.22}$ & 18.82 $\pm$ 0.01 & 0.16$_{-0.03}^{+0.01}$ & 68.56/75 & 0.76 $\pm$ 0.06 & 3.11 $\pm$ 0.05 & 1.13 $\pm$ 0.09 & 10.36 $\pm$ 0.05 & 10.94 $\pm$ 0.42 & \\ 

\rowcolor{gray!15}
12 & 51299.067 & 2.44$_{-0.05}^{+0.06}$ & 1.72$_{-0.08}^{+0.12}$ & 0.20$_{-0.03}^{+0.04}$ & 1.60$_{-0.02}^{+0.03}$ & 4.58$_{-2.08}^{+2.20}$ & 3.09 $\pm$ 0.22 & 18.86$_{-0.01}^{+0.01}$ & 0.04$_{-0.01}^{+0.01}$ & 67.96/73 & 1.25 $\pm$ 0.04 & 2.95 $\pm$ 0.04 & 0.55 $\pm$ 0.05 & 7.91 $\pm$ 0.05 & 13.88 $\pm$ 1.02 & \\ 

\rowcolor{gray!15}
13 & 51432.964 & 2.51$_{-0.11}^{+0.10}$ & 1.92$_{-0.04}^{+0.03}$ & 0.22$_{-0.04}^{+0.06}$ & 1.90$_{-0.05}^{+0.04}$ & 2.13$_{-0.85}^{+0.36}$ & 3.31$_{-0.09}^{+0.22}$ & 18.83 $\pm$ 0.01 & 0.12$_{-0.01}^{+0.02}$ & 57.00/69 & 0.82 $\pm$ 0.08 & 3.18 $\pm$ 0.08 & 1.41 $\pm$ 0.07 & 9.74 $\pm$ 0.07 & 11.59 $\pm$ 0.38 & \\ 

\rowcolor{gray!15}
14 & 51440.608 & 2.55$_{-0.11}^{+0.03}$ & 2.29$_{-0.13}^{+0.08}$ & 0.27$_{-0.03}^{+0.04}$ & 1.44$_{-0.02}^{+0.01}$ & 3.23$_{-1.99}^{+3.87}$ & 1.28$_{-0.65}^{+0.43}$ & 17.01$_{-1.67}^{+1.00}$ & 0.35 $\pm$ 0.13 & 76.51/72 & 1.36 $\pm$ 0.04 & 3.90 $\pm$ 0.04 & 0.17 $\pm$ 0.01 & 9.16 $\pm$ 0.04 & 9.03 $\pm$ 0.71 & \\ 

15 & 54285.950 & 3.20$_{-0.14}^{+0.04}$ & 3.00* & 0.64$_{-0.03}^{+0.08}$ & 1.57$_{-0.06}^{+0.02}$ & 4.00* & 2.91$_{-1.08}^{+0.26}$ & 16.94$_{-0.59}^{+1.03}$ & 0.71$_{-0.34}^{+1.17}$ & 64.29/65 & 1.34 $\pm$ 0.04 & 3.08 $\pm$ 0.04 & 0.22 $\pm$ 0.02 & 11.78 $\pm$ 0.16 & 5.58 $\pm$ 0.04 & \\ 

16 & 57509.889 & 3.13$_{-0.06}^{+0.07}$ & 2.40$_{-0.04}^{+0.05}$ & 0.68$_{-0.04}^{+0.02}$ & 1.86$_{-0.01}^{+0.03}$ & 12.87$_{-2.65}^{+5.58}$ & 4.46$_{-0.07}^{+0.09}$ & 17.67$_{-0.17}^{+0.16}$ & 0.08 $\pm$ 0.01 & 1347.90/1226 & 1.08 $\pm$ 0.01 & 1.82 $\pm$ 0.01 & 0.65 $\pm$ 0.01 & 6.08 $\pm$ 0.01 & 7.54 $\pm$ 0.18 & \\ 

17 & 58008.080 & 3.18$_{-0.11}^{+0.08}$ & 2.30$_{-0.03}^{+0.02}$ & 0.60$_{-0.04}^{+0.05}$ & 1.46$_{-0.04}^{+0.05}$ & 5.85$_{-2.72}^{+5.64}$ & 1.48$_{-0.07}^{+0.11}$ & 19.89$_{-0.19}^{+0.08}$ & 0.26$_{-0.02}^{+0.03}$ & 551.72/562 & 1.69 $\pm$ 0.10 & 2.07 $\pm$ 0.09 & 0.65 $\pm$ 0.01 & 7.72 $\pm$ 0.06 & 7.90 $\pm$ 0.21 & \\ 

18 & 58018.827 & 2.94$_{-0.04}^{+0.08}$ & 2.33$_{-0.04}^{+0.06}$ & 0.91 $\pm$ 0.04 & 1.59$_{-0.03}^{+0.04}$ & 13.70$_{-1.69}^{+1.61}$ & 3.92 $\pm$ 0.02 & 18.82 $\pm$ 0.01 & 0.06 $\pm$ 0.01 & 1512.40/1440 & 1.58 $\pm$ 0.04 & 1.55 $\pm$ 0.04 & 0.95 $\pm$ 0.02 & 6.64 $\pm$ 0.01 & 8.13 $\pm$ 0.19 & \\ 
	 
	 \hline
&&&&&&& \multicolumn{3}{c}{$\delta$ Class}    \\
      \hline
\rowcolor{gray!15}
19  & 50246.004$^{a}$ & 2.97$_{-0.14}^{+0.24}$ & 1.95$_{-0.08}^{+0.09}$ & 0.08$_{-0.02}^{+0.01}$ & 1.85$_{-0.03}^{+0.05}$ & 2.57$_{-0.68}^{+1.92}$ & 2.67 $\pm$ 0.01 & 19.00$_{-0.43}^{+0.46}$ & 0.10 $\pm$ 0.02 & 40.22/46 & 0.57 $\pm$ 0.01 & 4.82 $\pm$ 0.01 & 0.68 $\pm$ 0.01 & 10.41 $\pm$ 0.02 & 10.32 $\pm$ 0.66  \\

\rowcolor{gray!15}
20  & 50256.611$^{a}$ & 2.23$_{-0.05}^{+0.04}$ & 2.18$_{-0.02}^{+0.04}$ & 0.28$_{-0.04}^{+0.06}$ & 1.81 $\pm$ 0.03 & 2.55$_{-0.33}^{+0.99}$ & 2.40$_{-0.06}^{+0.09}$ & 18.38$_{-0.90}^{+0.18}$ & 0.36$_{-0.03}^{+0.05}$ & 33.89/46 & 0.49 $\pm$ 0.02 & 2.23 $\pm$ 0.02 & 1.00 $\pm$ 0.01 & 8.43 $\pm$ 0.02 & 10.40 $\pm$ 0.18  \\  
  
\rowcolor{gray!15}
21  & 50259.080$^{a}$ & 2.33$_{-0.09}^{+0.06}$ & 1.98 $\pm$ 0.03 & 0.20$_{-0.05}^{+0.08}$ & 1.86$_{-0.07}^{+0.06}$ & 3.25$_{-1.42}^{+1.33}$ & 2.68 $\pm$ 0.04 & 19.01$_{-0.29}^{+0.21}$ & 0.16$_{-0.01}^{+0.03}$ & 46.28/46 & 0.52 $\pm$ 0.01 & 2.50 $\pm$ 0.01 & 1.09 $\pm$ 0.01 & 8.13 $\pm$ 0.02 & 11.57 $\pm$ 0.35  \\ 

\rowcolor{gray!15}
22 & 50679.238 & 2.49$_{-0.07}^{+0.03}$ & 2.10$_{-0.09}^{+0.05}$ & 0.18$_{-0.07}^{+0.03}$ & 1.74$_{-0.03}^{+0.01}$ & 2.47$_{-0.91}^{+1.45}$ & 2.88$_{-0.08}^{+0.14}$ & 18.99$_{-0.73}^{+0.44}$ & 0.10 $\pm$ 0.01 & 81.27/77 & 0.82 $\pm$ 0.01 & 4.14 $\pm$ 0.01 & 0.72 $\pm$ 0.01 & 10.25 $\pm$ 0.01 & 10.27 $\pm$ 0.62 & \\ 

\rowcolor{gray!15}
23 & 50679.305 & 2.33$_{-0.07}^{+0.12}$ & 1.75$_{-0.10}^{+0.13}$ & 0.11 $\pm$ 0.02 & 1.68$_{-0.04}^{+0.03}$ & 2.50* & 4.00$_{-0.82}^{+0.46}$ & 17.73$_{-2.35}^{+1.26}$ & 0.04 $\pm$ 0.01 & 68.51/77 & 0.80 $\pm$ 0.04 & 3.97 $\pm$ 0.04 & 0.51 $\pm$ 0.05 & 8.70 $\pm$ 0.03 & 13.86 $\pm$ 1.23 & \\ 

\rowcolor{gray!15}
24 & 50679.463 & 2.44 $\pm$ 0.03 & 1.69$_{-0.14}^{+0.25}$ & 0.10$_{-0.04}^{+0.05}$ & 1.60 $\pm$ 0.02 & 2.67$_{-0.26}^{+0.24}$ & 3.00* & 15.00* & 0.03 $\pm$ 0.01 & 71.24/82 & 0.97 $\pm$ 0.01 & 4.49 $\pm$ 0.01 & 0.19 $\pm$ 0.01 & 9.36 $\pm$ 0.01 & 14.28 $\pm$ 1.88 & \\ 

\rowcolor{gray!15}
25 & 50681.796 & 2.49$_{-0.09}^{+0.11}$ & 2.07$_{-0.10}^{+0.16}$ & 0.17$_{-0.06}^{+0.07}$ & 1.73$_{-0.07}^{+0.14}$ & 2.42$_{-1.21}^{+1.97}$ & 2.86$_{-0.37}^{+0.27}$ & 18.86$_{-3.35}^{+1.01}$ & 0.100$_{-0.020}^{+0.010}$ & 68.24/76 & 0.83 $\pm$ 0.01 & 4.29 $\pm$ 0.01 & 0.70 $\pm$ 0.01 & 10.42 $\pm$ 0.01 & 10.48 $\pm$ 1.13 & \\ 

26 & 50763.209 & 2.37 $\pm$ 0.03 & 2.60* & 0.65$_{-0.05}^{+0.06}$ & 1.32 $\pm$ 0.04 & 4.00* & 1.28$_{-0.40}^{+0.22}$ & 17.00* & 1.50$_{-0.15}^{+0.07}$ & 85.26/81 & 2.00 $\pm$ 0.07 & 2.98 $\pm$ 0.07 & 0.26 $\pm$ 0.01 & 8.81 $\pm$ 0.05 & 7.99 $\pm$ 0.05 & \\ 

27 & 50763.229 & 3.00$_{-0.04}^{+0.18}$ & 2.80$_{-0.08}^{+0.01}$ & 0.67$_{-0.11}^{+0.03}$ & 1.61 $\pm$ 0.02 & 4.06$_{-1.56}^{+1.18}$ & 2.36$_{-0.05}^{+0.26}$ & 19.00$_{-0.02}^{+0.01}$ & 0.05 $\pm$ 0.01 & 84.81/74 & 1.18 $\pm$ 0.01 & 2.41 $\pm$ 0.01 & 0.48 $\pm$ 0.01 & 7.03 $\pm$ 0.01 & 6.33 $\pm$ 0.23 & \\ 

28 & 50774.223 & 2.31$_{-0.02}^{+0.03}$ & 2.61 $\pm$ 0.01 & 0.87$_{-0.06}^{+0.05}$ & 1.39 $\pm$ 0.02 & 4.02 $\pm$ 1.44 & 2.69 $\pm$ 0.01 & 19.47$_{-0.05}^{+0.10}$ & 0.024 $\pm$ 0.003 & 71.29/74 & 1.67 $\pm$ 0.06 & 2.13 $\pm$ 0.06 & 0.61 $\pm$ 0.01 & 7.52 $\pm$ 0.02 & 8.07 $\pm$ 0.05 & \\ 

29 & 50804.908 & 3.45$_{-0.26}^{+0.36}$ & 4.45$_{-0.30}^{+0.39}$ & 0.84$_{-0.11}^{+0.10}$ & 1.66$_{-0.02}^{+0.03}$ & 7.70$_{-1.97}^{+3.88}$ & 2.30$_{-0.01}^{+0.02}$ & 19.47$_{-0.06}^{+0.05}$ & 0.008$\pm$ 0.001 & 71.17/76 & 1.06 $\pm$ 0.02 & 4.63 $\pm$ 0.02 & 0.88 $\pm$ 0.01 & 10.69 $\pm$ 0.01 & 3.21 $\pm$ 0.32 & \\ 

30 & 52933.627 & 3.14$_{-0.08}^{+0.22}$ & 2.51$_{-0.06}^{+0.13}$ & 0.75$_{-0.03}^{+0.05}$ & 1.50$_{-0.03}^{+0.04}$ & 3.83$_{-2.51}^{+2.16}$ & 2.68$_{-0.27}^{+0.32}$ & 16.77$_{-1.24}^{+0.35}$ & 0.29$_{-0.07}^{+0.08}$ & 64.94/74 & 1.74 $\pm$ 0.01 & 2.12 $\pm$ 0.02 & 0.23 $\pm$ 0.01 & 7.01 $\pm$ 0.02 & 7.10 $\pm$ 0.56 & \\ 

31 & 52933.695 & 3.27$_{-0.21}^{+0.26}$ & 2.49$_{-0.15}^{+0.14}$ & 0.78$_{-0.09}^{+0.06}$ & 1.48$_{-0.04}^{+0.03}$ & 6.70$_{-2.69}^{+4.59}$ & 3.18$_{-0.36}^{+0.35}$ & 16.69$_{-1.50}^{+0.49}$ & 0.13$_{-0.06}^{+0.08}$ & 71.35/73 & 1.80 $\pm$ 0.02 & 2.01 $\pm$ 0.02 & 0.20 $\pm$ 0.01 & 6.64 $\pm$ 0.02 & 7.00 $\pm$ 0.61 & \\ 

32 & 52933.763 & 3.15$_{-0.08}^{+0.18}$ & 2.37$_{-0.09}^{+0.12}$ & 0.74$_{-0.07}^{+0.09}$ & 1.46 $\pm$ 0.02 & 4.76$_{-2.85}^{+2.96}$ & 2.29$_{-0.85}^{+0.18}$ & 18.23$_{-0.28}^{+1.41}$ & 0.15$_{-0.04}^{+0.03}$ & 68.62/73 & 1.80 $\pm$ 0.01 & 1.86 $\pm$ 0.03 & 0.21 $\pm$ 0.01 & 6.68 $\pm$ 0.04 & 7.64 $\pm$ 0.57 & \\ 

33 & 54326.708 & 3.03$_{-0.15}^{+0.08}$ & 2.66$_{-0.13}^{+0.05}$ & 0.77$_{-0.12}^{+0.06}$ & 1.22$_{-0.03}^{+0.02}$ & 5.38$_{-0.58}^{+1.22}$ & 1.69$_{-0.12}^{+0.20}$ & 20.00$_{-0.17}^{+p}$ & 0.04 $\pm$ 0.01 & 69.85/70 & 0.94 $\pm$ 0.03 & 0.93 $\pm$ 0.03 & 0.42 $\pm$ 0.01 & 5.15 $\pm$ 0.07 & 6.72 $\pm$ 0.19 & \\ 

34 & 57340.553 & 2.99$_{-0.06}^{+0.08}$ & 2.88$_{-0.05}^{+0.04}$ & 0.86$_{-0.04}^{+0.06}$ & 1.46$_{-0.04}^{+0.02}$ & 7.00* & 2.47$_{-0.07}^{+0.30}$ & 17.08$_{-0.21}^{+0.06}$ & 1.19$_{-0.15}^{+0.14}$ & 1256.18/1058 & 1.42 $\pm$ 0.01 & 2.08 $\pm$ 0.03 & 0.46 $\pm$ 0.02 & 10.11 $\pm$ 0.01 & 6.12 $\pm$ 0.01 & \\ 

35 & 57551.040 & 2.97 $\pm$ 0.06 & 2.40$_{-0.06}^{+0.03}$ & 0.76$_{-0.05}^{+0.06}$ & 1.07$_{-0.05}^{+0.04}$ & 3.74$_{-1.09}^{+2.36}$ & 1.99$_{-0.09}^{+0.22}$ & 18.89 $\pm$ 0.18 & 0.48$_{-0.13}^{+0.12}$ & 298.93/244 & 2.35 $\pm$ 0.24 & 1.37 $\pm$ 0.22 & 0.79 $\pm$ 0.02 & 9.01 $\pm$ 0.18 & 7.77 $\pm$ 0.27 & \\ 

36 & 57552.350 & 2.40$_{-0.06}^{+0.05}$ & 2.36$_{-0.05}^{+0.06}$ & 0.62$_{-0.08}^{+0.10}$ & 1.16$_{-0.07}^{+0.04}$ & 5.00* & 2.24$_{-0.17}^{+0.40}$ & 18.86$_{-0.44}^{+0.27}$ & 0.38$_{-0.10}^{+0.14}$ & 346.10/282 & 2.25 $\pm$ 0.20 & 1.96 $\pm$ 0.17 & 0.95 $\pm$ 0.02 & 10.42 $\pm$ 0.18 & 8.98 $\pm$ 0.27 & \\ 

\rowcolor{gray!15}
37 & 57552.866 & 2.20$_{-0.10}^{+0.06}$ & 2.04$_{-0.04}^{+0.03}$ & 0.33$_{-0.11}^{+0.16}$ & 1.86$_{-0.11}^{+0.10}$ & 3.00* & 3.25$_{-0.20}^{+0.21}$ & 18.89$_{-0.18}^{+0.09}$ & 0.15$_{-0.01}^{+0.02}$ & 560.94/472 & 0.93 $\pm$ 0.47 & 3.06 $\pm$ 0.46 & 1.44 $\pm$ 0.02 & 8.11 $\pm$ 0.08 & 11.46 $\pm$ 0.35 & \\ 

38 & 57553.074 & 2.93$_{-0.09}^{+0.12}$ & 3.01$_{-0.06}^{+0.23}$ & 0.84$_{-0.14}^{+0.14}$ & 1.56$_{-0.08}^{+0.16}$ & 8.27$_{-4.05}^{+2.10}$ & 2.41$_{-0.01}^{+0.55}$ & 16.88$_{-0.89}^{+0.01}$ & 2.830$_{-0.550}^{+3.110}$ & 1353.84/1176 & 1.38 $\pm$ 0.02 & 2.54 $\pm$ 0.02 & 0.55 $\pm$ 0.04 & 13.32 $\pm$ 0.03 & 5.86 $\pm$ 0.02 & \\

39 & 57553.880 & 2.61$_{-0.08}^{+0.06}$ & 2.40$_{-0.04}^{+0.02}$ & 0.90$_{-0.11}^{+0.08}$ & 1.44$_{-0.05}^{+0.08}$ & 5.77$_{-2.41}^{+2.69}$ & 1.78$_{-0.05}^{+0.09}$ & 18.95$_{-0.04}^{+0.09}$ & 0.99$_{-0.20}^{+0.08}$ & 424.99/426 & 2.17 $\pm$ 0.28 & 2.20 $\pm$ 0.27 & 0.70 $\pm$ 0.01 & 10.63 $\pm$ 0.15 & 8.37 $\pm$ 0.23 & \\ 

40 & 57554.090 & 2.90$_{-0.11}^{+0.14}$ & 2.38$_{-0.07}^{+0.09}$ & 0.71$_{-0.07}^{+0.11}$ & 1.10$_{-0.05}^{+0.04}$ & 3.87$_{-2.47}^{+5.67}$ & 2.01$_{-0.13}^{+0.19}$ & 18.80$_{-0.42}^{+0.21}$ & 0.52$_{-0.13}^{+0.23}$ & 138.94/143 & 2.54 $\pm$ 0.34 & 1.64 $\pm$ 0.30 & 0.77 $\pm$ 0.03 & 9.87 $\pm$ 0.32 & 7.97 $\pm$ 0.35 & \\ 

41 & 57588.595 & 3.47$_{-0.07}^{+0.07}$ & 3.09$_{-0.01}^{+0.03}$ & 0.90$_{-0.01}^{+0.09}$ & 1.33$_{-0.01}^{+0.02}$ & 3.16$_{-0.45}^{+0.38}$ & 1.30$_{-0.01}^{+0.03}$ & 18.84 $\pm$ 0.01 & 0.71$_{-0.07}^{+0.02}$ & 1316.66/1067 & 0.37 $\pm$ 0.04 & 0.45 $\pm$ 0.04 & 0.43 $\pm$ 0.02 & 3.40 $\pm$ 0.19 & 5.11 $\pm$ 0.06 & \\ 

42 & 57928.685 & 2.44$_{-0.16}^{+0.07}$ & 2.36$_{-0.06}^{+0.02}$ & 0.38$_{-0.10}^{+0.11}$ & 1.90$_{-0.08}^{+0.07}$ & 1.80$_{-0.26}^{+0.69}$ & 2.94$_{-0.13}^{+0.07}$ & 17.92$_{-0.12}^{+0.72}$ & 0.13$_{-0.04}^{+0.01}$ & 1477.57/1221 & 0.18 $\pm$ 0.02 & 0.68 $\pm$ 0.02 & 0.38 $\pm$ 0.01 & 2.84 $\pm$ 0.02 & 8.90 $\pm$ 0.32 & \\ 

43 & 58038.902 & 2.34$_{-0.06}^{+0.08}$ & 2.35$_{-0.03}^{+0.02}$ & 0.31$_{-0.09}^{+0.07}$ & 1.85$_{-0.06}^{+0.04}$ & 1.94$_{-0.22}^{+0.58}$ & 3.23$_{-0.10}^{+0.11}$ & 17.93$_{-0.12}^{+0.19}$ & 0.26$_{-0.05}^{+0.03}$ & 1472.67/1154 & 0.65 $\pm$ 0.02 & 2.87 $\pm$ 0.02 & 1.23 $\pm$ 0.01 & 10.63 $\pm$ 0.02 & 9.16 $\pm$ 0.20 & \\ 

\rowcolor{gray!15}
44 & 58045.931 & 2.38$_{-0.16}^{+0.02}$ & 1.85$_{-0.11}^{+0.05}$ & 0.20$_{-0.03}^{+0.04}$ & 1.15$_{-0.05}^{+0.04}$ & 2.09$_{-1.86}^{+0.51}$ & 3.07$_{-0.31}^{+0.55}$ & 18.87$_{-0.01}^{+0.03}$ & 0.04 $\pm$ 0.01 & 300.46/253 & 0.62 $\pm$ 0.07 & 0.97 $\pm$ 0.05 & 0.32 $\pm$ 0.02 & 3.35 $\pm$ 0.07 & 12.60 $\pm$ 1.19 & \\ 

 \hline

&&&&&&& \multicolumn{3}{c}{$\phi$ Class}    \\
      \hline

\rowcolor{gray!15}
45 & 50232.530$^{a}$ & 2.28 $\pm$ 0.07 & 1.82$_{-0.23}^{+0.19}$ & 0.25$_{-0.10}^{+0.13}$ & 1.23$_{-0.09}^{+0.05}$ & 2.50* & 3.62* & 18.82* & 0.014 $\pm$ 0.004 & 59.25/49 & 0.82 $\pm$ 0.01 & 1.16 $\pm$ 0.01 & 0.20 $\pm$ 0.01 & 3.94 $\pm$ 0.01 & 13.22 $\pm$ 2.49 & \\ 

\rowcolor{gray!15}
46 & 50239.483$^{a}$ & 2.03$_{-0.05}^{+0.04}$ & 1.60* & 0.16$_{0.02}^{+0.03}$ & 1.37$_{-0.05}^{+0.03}$ & 2.50* & 3.79* & 18.05* & 0.016$_{-0.001}^{+0.002}$ & 49.73/50 & 0.71 $\pm$ 0.01 & 1.46 $\pm$ 0.01 & 0.28 $\pm$ 0.01 & 4.25 $\pm$ 0.01 & 17.25 $\pm$ 0.22 & \\ 

\rowcolor{gray!15}
47 & 50256.744$^{a}$ & 2.50* & 2.28 $\pm$ 0.02 & 0.11$_{-0.01}^{+0.06}$ & 1.82$_{-0.11}^{+0.02}$ & 2.17$_{-0.22}^{+0.76}$ & 1.88$_{-0.05}^{+0.41}$ & 19.40$_{-1.02}^{+0.07}$ & 0.20$_{-0.01}^{+0.02}$ & 70.71/74 & 0.21 $\pm$ 0.01 & 1.98 $\pm$ 0.01 & 0.67 $\pm$ 0.01 & 6.43 $\pm$ 0.01 & 9.18 $\pm$ 0.10 & \\ 

\rowcolor{gray!15}
48 & 50263.550$^{a}$ & 2.18$_{-0.09}^{+0.21}$ & 1.84$_{-0.17}^{+0.25}$ & 0.16$_{-0.04}^{+0.05}$ & 1.22$_{-0.01}^{+0.02}$ & 2.50* & 3.76$_{-0.94}^{+0.76}$ & 17.90$_{-2.61}^{+0.77}$ & 0.007 $\pm$ 0.002 & 47.51/48 & 0.33 $\pm$ 0.01 & 0.78 $\pm$ 0.01 & 0.08 $\pm$ 0.01 & 2.28 $\pm$ 0.01 & 13.34 $\pm$ 1.83 & \\ 

\rowcolor{gray!15}
49 & 50267.486$^{a}$ & 2.09$_{-0.04}^{+0.09}$ & 1.58$_{-0.05}^{+0.09}$ & 0.11$_{-0.01}^{+0.02}$ & 1.30$_{-0.02}^{+0.03}$ & 2.50* & 4.21$_{-0.29}^{+0.42}$ & 17.98$_{-1.32}^{+0.88}$ & 0.010$_{-0.001}^{+0.002}$ & 52.43/47 & 0.41 $\pm$ 0.01 & 1.06 $\pm$ 0.01 & 0.21 $\pm$ 0.01 & 3.06 $\pm$ 0.01 & 17.35 $\pm$ 0.95 & \\ 

\rowcolor{gray!15}
50 & 54326.316 & 2.08$_{-0.03}^{+0.01}$ & 2.39 $\pm$ 0.01 & 0.23$_{-0.01}^{+0.02}$ & 1.14$_{-0.01}^{+0.03}$ & 3.00$_{-0.02}^{+0.01}$ & 4.70$_{-0.05}^{+p}$ & 18.62$_{-0.03}^{+0.01}$ & 0.009 $\pm$ 0.001 & 78.13/73 & 0.27 $\pm$ 0.01 & 0.77 $\pm$ 0.01 & 0.33 $\pm$ 0.01 & 2.79 $\pm$ 0.01 & 9.59 $\pm$ 0.09 & \\

      \hline
  
	 \end{tabular}
}	 

	 \begin{list}{}{}
            \item[a] The \textit{PCA} spectra are used only for these observations due to the unavailability of \textit{HEXTE} background.
		\item[*] Parameter is fixed.
            \item[p] Parameter is pegged at the higher or lower limit. 
            \item[$\dagger$] The fluxes for model components are calculated in the energy range $3-50$ keV in the unit of $10^{-8}$ erg cm$^{-2}$ s$^{-1}$.
            \item[$\ddagger$] The bolometric flux is calculated in the energy range $1-100$ keV in the unit of $10^{-8}$ erg cm$^{-2}$ s$^{-1}$.
	\end{list}

\end{table}


\end{landscape}

Fig. \ref{fig:spec_correlation0} demonstrates a quantitative limitation in covering fraction ($f_{cov} \gtrsim 0.5$) and optical depth ($\tau \lesssim 8.5$ ) values that indicates when HFQPOs are likely to be observable. When HFQPO is observed, $f_{cov}$ increases while $\tau$ decreases. Conversely, during periods without HFQPO, $f_{cov}$ is lower and $\tau$ is higher.
Most of the HFQPO observations show a larger inner disc radius ($R_{in}$) compared to observations without HFQPO, indicating that the accretion disc is located farther from the black hole when HFQPOs are present. Additionally, both $f_{cov}$ and the average percentage of Comptonized flux are higher when HFQPOs are likely to be observed.
The findings shown in Fig. \ref{fig:spec_correlation0}, impose constraints on spectral parameters, such as covering fraction ($f_{cov}$), optical depth ($\tau$), and inner disc radius ($R_{in}$), for the likelihood of detecting HFQPOs.

\subsection{ Time-lag and Optical Depth Correlation }

\begin{figure}
 \includegraphics[width=\columnwidth]{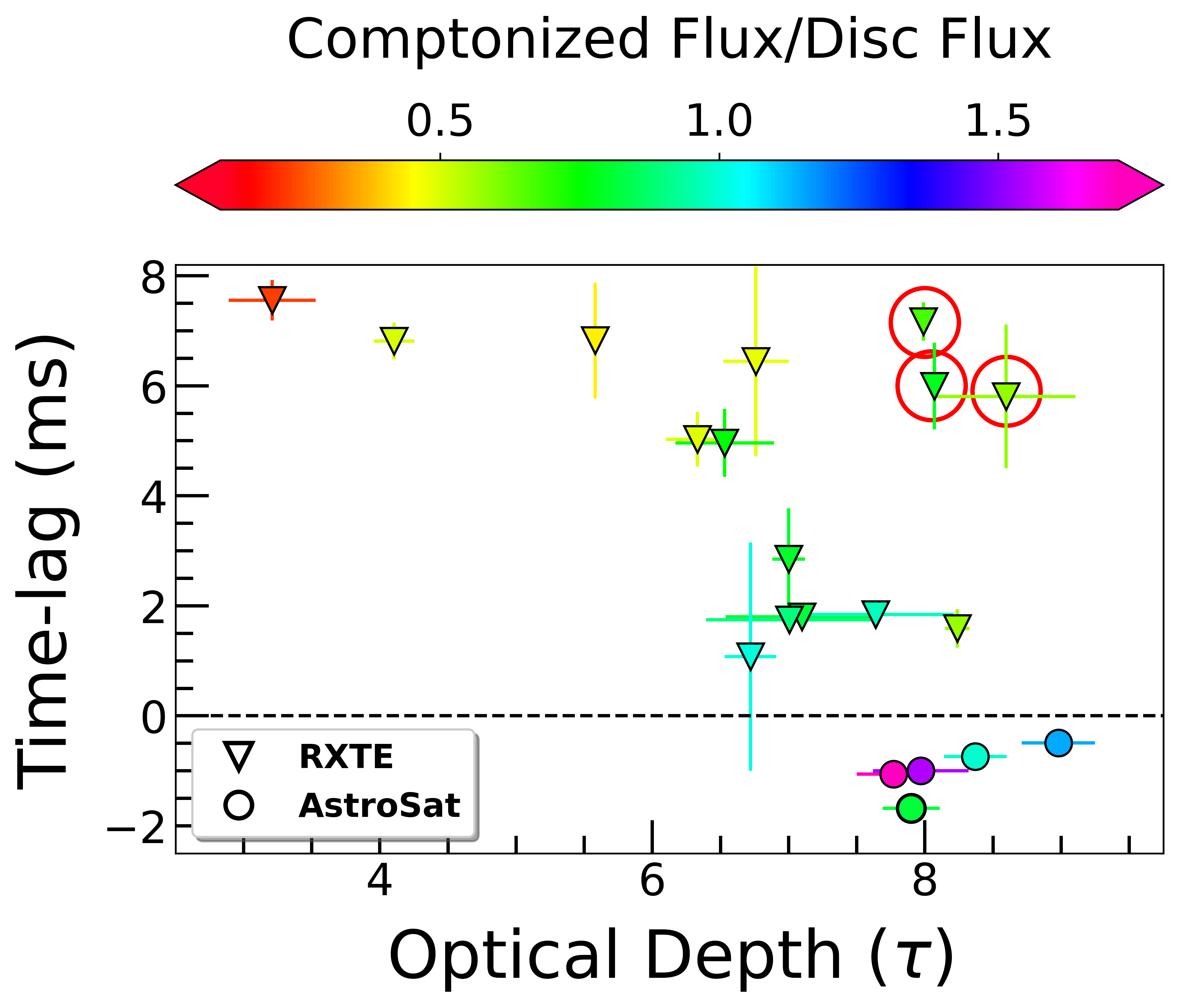}
 \caption{The variation of the time-lag is plotted with optical depth along with the ratio of Comptonized flux and disc flux as the colour bar. The mission of the observations are shown in the legend. The outliers are marked in red circles. See text for details.}
 \label{fig:spectro_temporal2}
\end{figure}

The time-lag study at HFQPO shows hard-lag of $5-25$ keV photons w.r.t $2-5$ keV photons in \textit{RXTE} observations whereas, soft-lag is found for $6-25$ keV photons w.r.t $3-6$ keV photons in \textit{AstroSat} observations (see \S \ref{subsec:time-lag}). The spectral analysis suggests a correlation between the optical depth ($\tau$) and the time-lag of hard photons ($5-25$ keV or $6-25$ keV) w.r.t soft photons ($2-5$ keV or $3-6$ keV) (see Table \ref{tab:pds_table}). 
The variation of the time-lag in milliseconds is plotted as a function of the optical depth of the corona in Fig. \ref{fig:spectro_temporal2}. The \textit{RXTE} and \textit{AstroSat} observations are indicated by filled triangles and circles respectively and the colour of each data point reflects the value of the Comptonization ratio (the ratio of Comptonized flux to disc flux). 

\begin{figure}
	\includegraphics[width=\columnwidth]{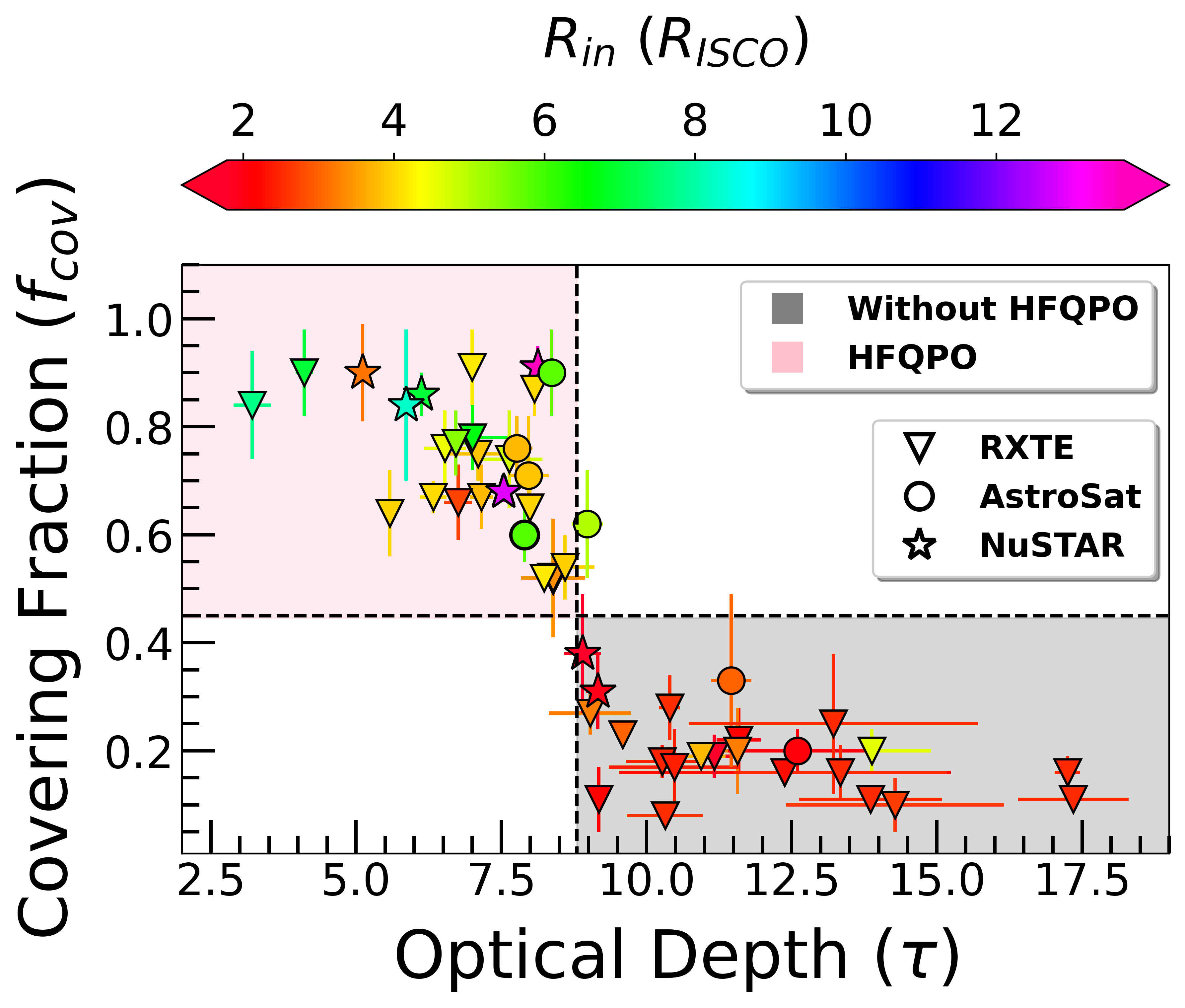}
    \caption{Correlation between optical depth ($\tau$) and covering fraction ($f_{cov}$) is plotted along with inner disc radius R$_{in}$ (R$_{ISCO}$) as colour bar. The grey and pink regions indicate the region of the observations without HFQPO and with HFQPO respectively. The mission of the observations are shown in the legend. See text for details.}
    \label{fig:spec_correlation0}
\end{figure}

As evident from Fig. \ref{fig:spectro_temporal2}, the time lag at HFQPO decreases with increasing $\tau$, eventually shifting from a hard-lag (positive) to a soft-lag (negative) from the \textit{RXTE} to \textit{AstroSat} observations as $\tau$ increases further. Furthermore, as $\tau$ increases further, HFQPOs disappear from the PDS (see Fig. \ref{fig:spec_correlation0}) imposing a constraint on $\tau$ for their detection. Therefore, an increase in optical depth leads to reduced Comptonization, resulting in shorter time delays for energized photons, i.e., a decrease in hard-lag.    
Initially, the Comptonization ratio is observed to increase as the time-lag decreases. 
However, we identified three outliers (highlighted as red circles in Fig. \ref{fig:spectro_temporal2}), where the time lag does not decrease with the optical depth of the Comptonized medium. However, as shown in Fig. \ref{fig:spectro_temporal2}, we find an overlapping optical depth range ($\sim$ $7.5 - 8.5$) where \textit{RXTE} observations show a positive lag, while \textit{AstroSat} observations show a negative lag, both with a small magnitude of time lag of $\sim 2 \pm 0.9 $ ms. This unusual variation is observed in only two out of 12 \textit{RXTE} observations (excluding outliers). We also expect that the observed lag variations cannot be explained solely by a single mechanism, as proposed in our previous studies \citep[][]{dutta2016,arka-chatterjee2020,patra2019,prajjwal_2024}. However, in this study, we find that optical depth variation could be a key parameter in explaining time lag variations from the \textit{RXTE} to \textit{AstroSat} era, which we will discuss in the next section.

\section{Discussion}
\label{sec:discussion}

In this work, we conducted an in-depth analysis of all `canonical' soft state observations of GRS 1915+105 spanning from 1996 to 2017 using data from \textit{RXTE}, \textit{AstroSat} and \textit{NuSTAR}. This work aimed to relate spectro-temporal parameters to investigate the generation of HFQPOs and the possible accretion dynamics during the `canonical' soft state.
As evident in Fig. \ref{fig:asm_maxi}, the long-term light curve of the source, observed using \textit{RXTE/ASM} and \textit{MAXI/GSC}, exhibits its persistent nature from 1996 to 2019 \citep{motta2021,athulya2022,athulya2023}.

\subsection{HFQPOs in `Canonical' Soft State}
\label{subsec:soft_grs}
Source variation during $\gamma$, $\delta$ and $\phi$ variability classes can be interpreted through the lightcurve and CCD analysis (see \S \ref{subsec:lightcurve_ccd}). 
These class variations can be attributed to the transitions between two spectral states: soft outburst (B) with higher count rates and HR1, and flaring state (A) with lower count rates, HR1, and HR2 \citep{belloni2000}.
The soft-colour (HR1 $\sim$ $0.49-1.30$) and hard-colour (HR2 $\sim$ $0.01-0.09$) confirm that the considered observations correspond to the soft state (see Table \ref{tab:log_table}).
The $\phi$ class is the softest among the three, corresponding to spectral state A \citep[see][]{belloni2000}. In this class, HR1, HR2, and Total$_{rms}$\% increase with the count rate (see Table \ref{tab:log_table} and \ref{tab:pds_table}). Interestingly, HFQPO features are absent in the PDS for $\phi$ class observations (see right panel of Fig. \ref{fig:lc_ccd_pds}).

HFQPOs are observed in the PDS of soft state observations only for the $\gamma$ and $\delta$ classes, with frequencies ranging from 65.07 to 71.38 Hz (rms $\sim$ 0.38\% $-$ 2.14\%; see Table \ref{tab:pds_table}). The total broadband (0.1$-$500 Hz) percentage rms across all observations lies between 5.04\% to 20.87\%.
HFQPOs are associated with state B, which is disc-dominated \citep[see][]{belloni2013}. 
Their absence in the $\phi$ class is expected, as it corresponds solely to state A \citep{belloni2000}. The $\gamma$ and $\delta$ classes, which contain both state A and B, typically exhibit HFQPOs \citep{belloni2000, belloni2013}. However, not all observations from these classes show HFQPOs, even when state B is present (see Table \ref{tab:pds_table}).

The source exhibits dual time-lag behaviour in the $\gamma$ and $\delta$ variability classes (see \S \ref{subsec:time-lag}) as shown in Fig. \ref{fig:ene_dep_time-lag}. The significant hard-lag associated with HFQPO was attributed to the presence of a `Compton corona’ that reprocesses the disc photons and yields HFQPO features \citep{remillard_2002,cui1999}.
We also find that the hard-lag increases with energy for both $\gamma$ and $\delta$ class observations in \textit{RXTE}, consistent with the findings of \cite{cui1999} and \cite{mendez2013} for other \textit{RXTE} observations. This further confirms the Comptonization mechanism in the corona. 
In a previous study, we detected and explained, for the first time,  an increase in soft-lag with energy during \textit{AstroSat} observations, likely due to the dominance of the reflection mechanism \citep[see for detail,][]{prajjwal_2024}. We verified this by obtaining similar results using the \texttt{LAXPCsoftware}, consistent with those reported by \cite{belloni2019} using the \texttt{GHATS} package.

The wide-band spectra (0.7$-$50 keV) of all soft state observations were fitted using a multi-coloured accretion disc model combined with a thermal Comptonization model, accounting for high-energy photons produced via inverse Compton scattering of soft seed photons from the disc.
A reflection feature is also observed in the \textit{NuSTAR} spectra (see \S \ref{subsec:spectral_analysis}). 
We observe a higher inner disc temperature of $\sim 1.5$ keV and a lower electron temperature of $\sim 3$ keV and also a higher bolometric flux (up to 75\% $L_{Edd}$). 
The variation in inner disc radius ($\sim$ 1.78 $-$ 13.70 $R_{ISCO}$) indicates the disc extended near the black hole during the soft state, resulting in higher disc temperatures.
In the `canonical' soft state, higher disc accretion rates extend the disc to the ISCO \citep[see][]{skc_dutta_pal_2009,dutta2016,belloni_review_2016}, cooling the corona and lowering the electron temperature compared to the hard state \citep{malzac_2012}.
A similar accretion scenario is seen in 4U 1543-47, with a high inner disc temperature (0.9$-$1.27 keV) and super-Eddington bolometric flux during its 2021 soft state outburst \citep{geethu_2023}.

\subsection{HFQPOs in Class Transition}
\label{subsec:transition_of_hfqpo}

We find similar variations of spectral parameters (see for detail \S \ref{subsec:spectro-temp-para}), as obtained from Fig. \ref{fig:spec_correlation0} during an inter-class ($\phi \rightarrow \delta$) transition and in a few intra-class ($\delta \rightarrow \delta$) variations, where the HFQPO appears, disappears, and reappears within a timescale of a few hours or days. 
During the $\delta$ variability class (from MJD 57551 to 57554) observed by \textit{AstroSat} (see Table \ref{tab:log_table}), the HFQPO was present in Obs No. 35 and 36, disappeared in Obs No. 37 after 12 hours, and reappeared in Obs No. 39 the following day.
Similarly, during the transition from the $\phi$ to $\delta$ class (from MJD 54326.316 (Obs 50) to 54326.708 (Obs 33)) as observed by \textit{RXTE}, HFQPO appeared within $\sim$ 9 hours of the class transition \citep{ueda_2009}.

\begin{figure*}
 \includegraphics[width=0.9\textwidth]{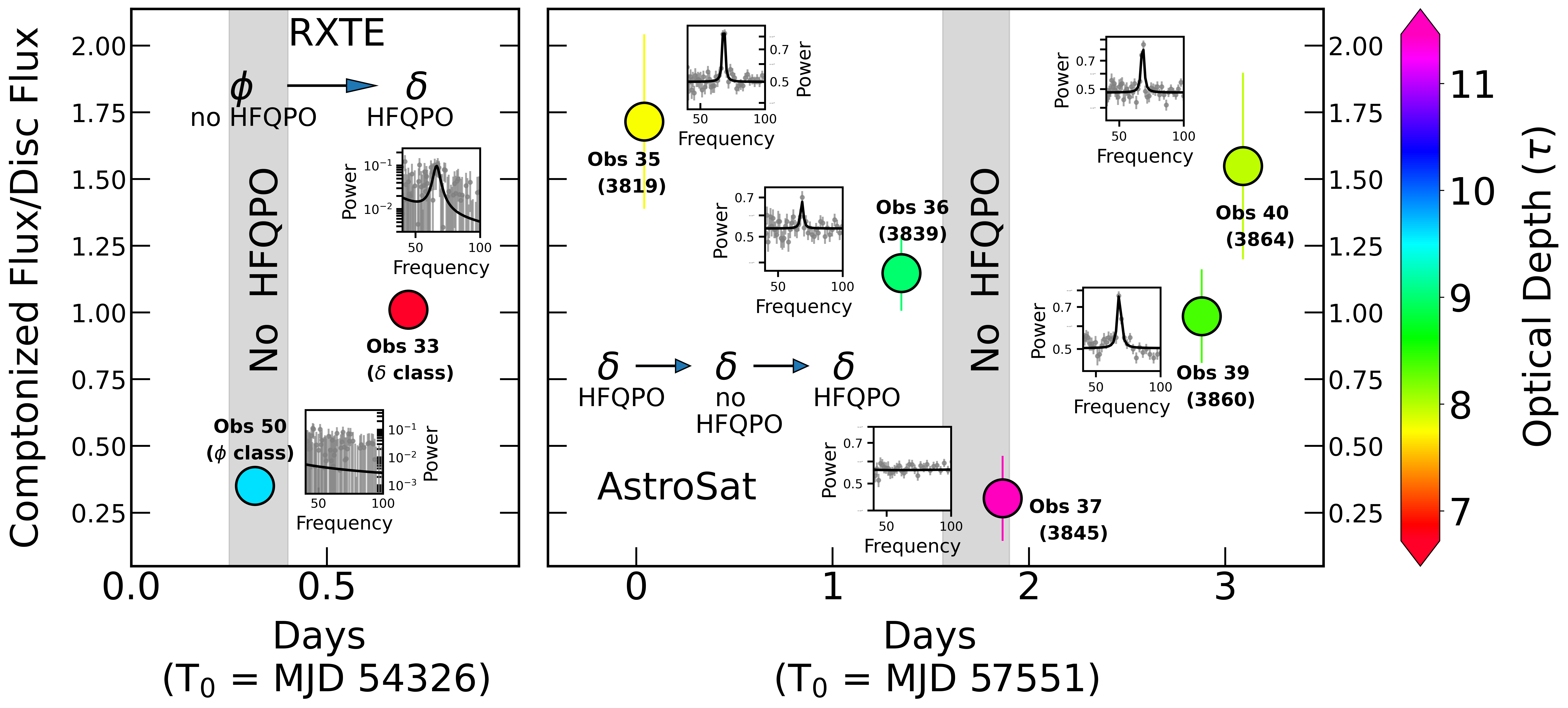}
 \caption{During inter-class transition and intra-class variation, the presence and absence of HFQPOs are shown along with the variation of the Comptonization ratio and optical depth (colour bar). The grey region represents the duration when HFQPOs are absent. For all observations, the PDS in the high-frequency domain is displayed in the inset.}
 \label{fig:transition}
\end{figure*}

Our broad-band spectral analysis indicates that HFQPO observations are associated with lower disc flux and higher Comptonization at lower optical depths ($\tau$), as shown in Fig. \ref{fig:spec_correlation0} during these transitions.
These results are qualitatively and quantitatively demonstrated in Fig. \ref{fig:transition} where optical depth gradually increases when HFQPOs disappear and decreases when HFQPOs reappear and a reverse mechanism for the Comptonization ratio is observed for observing HFQPOs in both inter-class and intra-class transitions. We find lower optical depth and higher Comptonization during HFQPO observations and the opposite is for no HFQPO observations which is also shown in Fig. \ref{fig:spec_correlation0}. This indicates that optical depth significantly influences the Comptonization flux and the generation of HFQPOs.

\subsection{Possible Accretion Dynamics in Soft State}
Several attempts have been made to explain the origin of HFQPOs in BH-XRBs through various accretion scenarios.
\cite{morgan1997} proposed that HFQPOs are associated with the Keplerian frequency of hot gas motion at the ISCO, but this model requires an unusually high source mass of $\sim$ 30 M$_\odot$. \cite{cui1999} and \cite{remillard_2002} suggested that hard-lags associated with HFQPOs arise from disc photons gaining energy through Compton upscattering in the `Compton corona', resulting in the HFQPO features.
In the framework of shock dominated two-component advective flow model \citep{chakrabarti_titarchuk1995},  the oscillations in the post-shock corona could produce modulated Comptonized radiation \citep{aktar2017,aktar2018}.
A satisfactory explanation of the observational features in sources like GRO J1655-40 and GRS 1915+105 \citep{dihingia2019,seshadri2022} within this model requires the presence of an oscillating corona in generating HFQPOs.

\begin{figure}
 \includegraphics[width=\columnwidth]{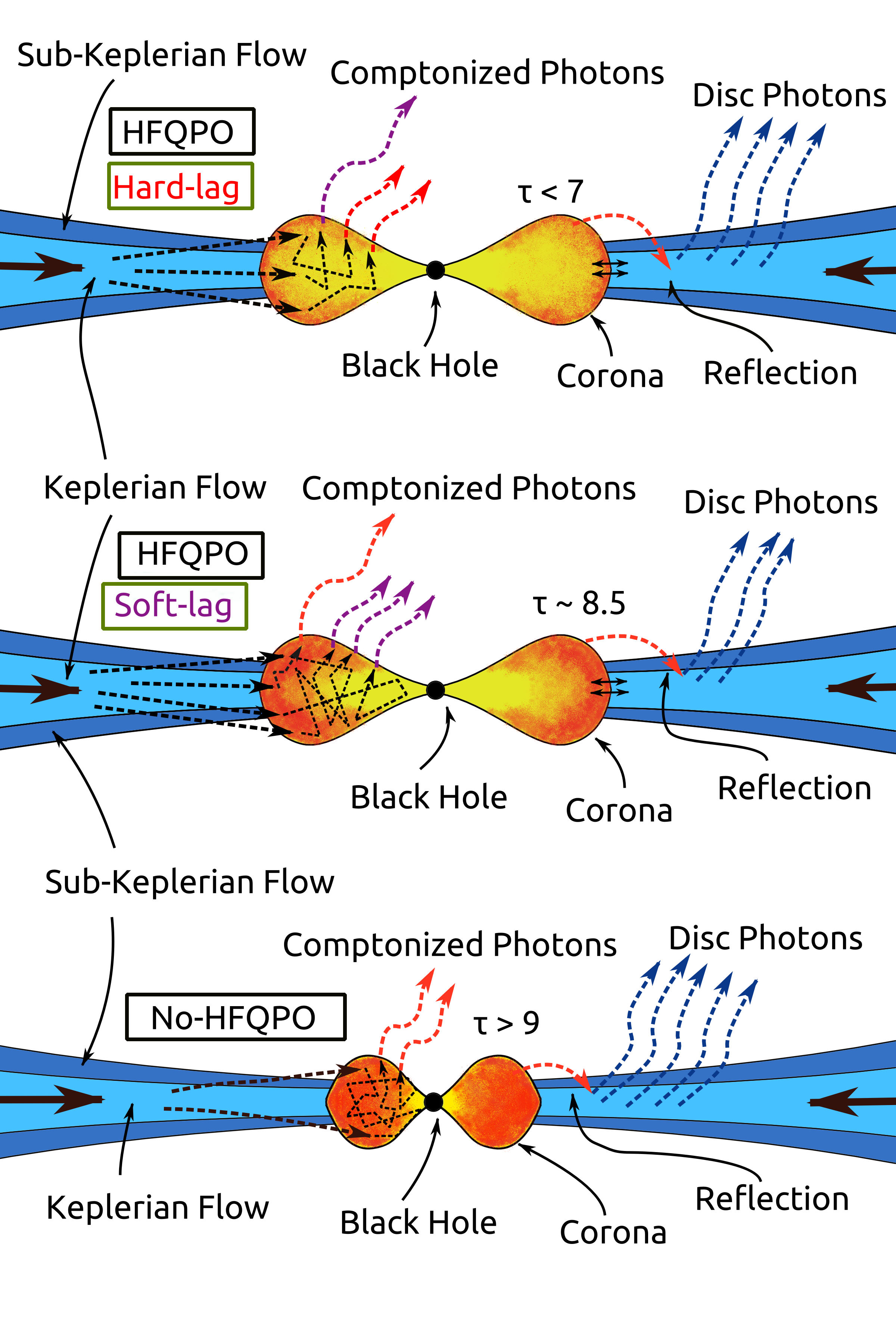}
 \caption{The schematic diagram depicting the accretion dynamics of the black hole source's HFQPOs shows hard-lag in the upper panel and soft-lag in the middle panel. The lower panel represent the accretion geometry in the absence of HFQPO. See text for details.}
 \label{fig:cartoon_diagram}
\end{figure}

To understand the generation of HFQPOs and possible accretion dynamics during the `canonical' soft state, we present a schematic diagram in Fig. \ref{fig:cartoon_diagram}. This depiction is derived from our spectro-temporal analysis and based on a two-component advective flow solution \citep{chakrabarti_titarchuk1995} and insights from our previous studies \citep[see details in][]{skc_dutta_pal_2009,dutta2016,arka-chatterjee2020,prajjwal_2024}. 
Our spectro-temporal analysis reveals that reduced disc flux, enhanced Comptonization with lower optical depth ($\tau$), are associated with the generation of HFQPOs (see \S \ref{subsec:spectro-temp-para}), as shown in Fig. \ref{fig:spec_correlation0}.

The top and middle panels of Fig. \ref{fig:cartoon_diagram} depict the possible origins of hard-lag and soft-lag in HFQPOs, along with $\tau$ variations.
The purple and red curved arrows in the upper and middle panels represent softer and harder photons emitted from the corona, respectively.
Figure \ref{fig:spectro_temporal2} shows that the time lag at HFQPO decreases with increasing $\tau$, and changes from a hard-lag to a soft-lag. This implies an increase in $\tau$ causes to less Comptonization, resulting in shorter time delays for energized photons, i.e., a decrease in hard-lag.
A corona with lower optical depth up-scatters lesser seed photons (black dashed arrows) from the accretion disc, resulting in a hard lag, as shown in the top panel of Fig. \ref{fig:cartoon_diagram}. Hard lag occurs when hard photons (red dashed arrows) lag behind soft photons (purple dashed arrows).
Consequently, if the corona's optical depth increases further, seed photons initially gain energy through up-scattering (producing hard photons) but lose energy through repeated down-scatterings (producing soft photons). In this scenario, soft photons are expected to lag behind hard photons \citep{reig2000}, as depicted in the middle panel of Fig. \ref{fig:cartoon_diagram}.

The lower panel of Fig. \ref{fig:cartoon_diagram} illustrates the accretion dynamics scenario in the absence of HFQPOs. Here, the inner disc radius moves closer to the black hole, increasing the average disc flux from 50.88\% to 64.92\%, while the disc temperature remains steady at $\sim$ 1.5 keV. A higher optical depth indicates a more `compact' corona with reduced Comptonization. We find that HFQPOs disappear from the PDS as the averaged Comptonized flux decreases from 37.39\% to 19.47\%, indicating a direct correlation with Comptonized photons. Hence, we conclude that  HFQPOs may originate from oscillations in the corona, as suggested by \cite{sreehari2020} and \cite{seshadri2022}.
Fig. \ref{fig:spec_correlation0} clearly differentiates the presence or absence of HFQPOs with covering fraction ($f_{cov}$) and optical 
depth($\tau$) values. When HFQPO is observed, $f_{cov}$ increases while $\tau$ decreases. Conversely, $f_{cov}$ is lower and $\tau$ is higher during periods without HFQPO. A similar result was found in the case of NGC 1566 where the optical depth of the corona is observed to decrease as the covering fraction increases \citep{p_tripathi_gulab_2022}. Moreover, most of our HFQPO observations show a larger inner disc radius ($R_{in}$) compared to non-HFQPO observations, indicating that the accretion disc is more distant when HFQPO is present.
Furthermore, our analysis shows that if the optical depth exceeds a threshold ($\tau \sim 8.8$), the electron temperature becomes so less ($2.03 - 2.55$ keV) that they are unable to Comptonize the seed photons at all. As a result, the covering fraction decreases, and the HFQPOs disappear. 
However, variations in the optical depth of the corona may be driven by changes in the accretion rate, jet activity, or outflows, leading to variations in the coronal radius and overall accretion geometry. During the spectral evolution of the source, once the soft state is reached, it remains in this state until the accretion rate declines to a few percent of the Eddington rate, eventually transitioning back to the hard state.
GRS 1915+105 has been accreting near the Eddington rate for more than a decade \citep{fender_belloni_2004}, powerful winds that carry away mass are observed in the softer X-ray states \citep{neilsen_2009_grs1915,ponti_2012} which may contribute to variations in the optical depth of the corona. The role of optical depth variation can be clearly understood from Fig. \ref{fig:transition}, where we observe the transition of HFQPOs (i.e., HFQPO $\rightarrow$ No HFQPO $\rightarrow$ HFQPO) during a long observation. During this intra-class transition (over a few days), the optical depth gradually increases when HFQPOs disappear and decreases when they reappear. This indicates that HFQPOs are mainly driven by variations in optical depth and Comptonized flux.

\section{Conclusion}
\label{sec:conclusion}
Understanding and interpreting the mechanism behind the generation of HFQPOs remains challenging and is still a topic of active debate. We performed a comprehensive analysis of all `canonical’ soft states of GRS 1915+105 from 1996 to 2017 using data from \textit{RXTE}, \textit{AstroSat}, and \textit{NuSTAR}, which often show HFQPOs in the PDS. The time lag at HFQPO exhibits an evolution from the \textit{RXTE} to \textit{AstroSat} era, possibly due to changes in the optical depth of the corona. Wide-band spectral study suggest that the Comptonization process is dominant when HFQPOs are present and fluctuating `compact' corona may be the source of HFQPOs \citep{sreehari2020,seshadri2022}. A summary of our findings is presented below.

\begin{itemize}
 \item HFQPO is observed only in $\gamma$ and $\delta$ class and the time-lag at HFQPO evolves from a hard-lag in \textit{RXTE} observation to a soft-lag in \textit{AstroSat} observation. These variations could be due to changes in optical depth.
 \item HFQPOs are more likely to be observed with a higher covering fraction, enhanced Comptonized flux, a more distant inner disc radius, and lower optical depth compared to when they are absent. 
 
 \item These correlated findings provide quantitative constraints on the spectral parameters that influence the likelihood of observing HFQPOs.
 \item The time lag at HFQPO decreases with increasing $\tau$, indicating that higher optical depth reduces Compton up-scattering, thereby decreasing the hard-lag.
\item During both intra-class variations and inter-class transitions, HFQPOs disappear as the $\tau$ increases and reappear as it decreases. Conversely, the Comptonization flux exhibits an inverse correlation with HFQPO generation, where higher $\tau$ results in reduced Comptonization flux and a lower likelihood of detecting HFQPOs.
   
\end{itemize}

\section*{Acknowledgments}
We thank the anonymous reviewer for valuable suggestions and comments that helped to improve the quality of this manuscript.
BGD, PM and AN acknowledge the support from ISRO sponsored project (DS-2B-1313(2)/6/2020-Sec.2). PM, BGD thanks the Department of Physics, Rishi Bankim Chandra College for providing the facilities to support this work. BGD acknowledges `TARE’ scheme (Ref. No. TAR/2020/000141) under SERB, DST, Govt. of India and also acknowledges Inter-University Centre for Astronomy and Astrophysics (IUCAA) for the Visiting Associate-ship Programme. AN thanks GH, SAG; DD, PDMSA, and Director, URSC for encouragement and continuous support to carry out this research. This work uses the data of \textit{AstroSat} mission of ISRO which is archived at the Indian Space Science Data Centre (ISSDC). We thank the SXT-POC team at TIFR for providing the necessary software tool to analyse \textit{SXT} data. This work has also used \textit{LAXPC} data which is verified by LAXPC-POC at TIFR. We thank \textit{AstroSat} Science Support Cell for providing the software \texttt{LAXPCsoftware} for the analysis of \textit{LAXPC} data. This publication made use of data from the \textit{RXTE} and \textit{NuSTAR} mission by the NASA. This work has also used data from Monitor of All-sky X-ray Image (\textit{MAXI}) data provided by Institute of Physical and Chemical Research (RIKEN), Japan Aerospace Exploration Agency (JAXA), and the \textit{MAXI} team. We also thank the High Energy Astrophysics Science Archive Research Center (HEASARC) team for providing the necessary software to analyse the data.

\section*{Data Availability}
Observational data of \textit{RXTE} and \textit{NuSTAR} used for this work are available at the HEASARC website  
\url{https://heasarc.gsfc.nasa.gov/cgi-bin/W3Browse/w3browse.pl}.
\textit{AstroSat} archival data used for this work is available at the Astrobrowse (AstroSat archive) website  
\url{https://webapps.issdc.gov.in/astro_archive/archive} of the Indian Space Science Data Centre (ISSDC). 
The \textit{MAXI/GSC} data is available in the website 
\url{http://maxi.riken.jp/top/lc.html}


\bibliographystyle{mnras}
\bibliography{reference} 







\bsp	
\label{lastpage}
\end{document}